\documentclass[sigconf]{acmart}

\usepackage{enumitem}
\usepackage{multirow}
\usepackage{array}
\newcommand{\hl}[1]{\textcolor{black}{#1}}

\newcommand{\xsc}[1]{\textcolor{black}{#1}}

\AtBeginDocument{%
  }

\copyrightyear{2025}
\acmYear{2025}
\setcopyright{acmlicensed}
\acmConference[UIST '25]{The 38th Annual ACM
Symposium on User Interface Software and Technology}{September 28-October
1, 2025}{Busan, Republic of Korea}
\acmBooktitle{The 38th Annual ACM Symposium on User Interface Software
and Technology (UIST '25), September 28-October 1, 2025, Busan, Republic of Korea}
\acmDOI{10.1145/3746059.3747791}
\acmISBN{979-8-4007-2037-6/2025/09}

\begin{document}

\title[Branch Explorer: Interactive 360° Video Viewing for Blind and Low Vision Users]{Branch Explorer: Leveraging Branching Narratives to Support Interactive 360° Video Viewing for Blind and Low Vision Users}

\author{Shuchang Xu}
\affiliation{
\institution{The Hong Kong University of Science and Technology}
\city{Hong Kong}
\country{China}
}
\orcid{0000-0002-7642-9044}
\email{sxuby@connect.ust.hk}

\author{Xiaofu Jin}
\affiliation{
\institution{The Hong Kong University of Science and Technology}
\city{Hong Kong}
\country{China}
}
\orcid{0000-0002-7239-3769}
\email{xjinao@connect.ust.hk}

\author{Wenshuo Zhang}
\affiliation{
\institution{The Hong Kong University of Science and Technology}
\city{Hong Kong}
\country{China}
}
\orcid{0009-0007-9226-0713}
\email{wzhangeb@connect.ust.hk}

\author{Huamin Qu}
\affiliation{
\institution{The  Hong Kong University of Science and Technology}
\city{Hong Kong}
\country{China}
}
\orcid{0000-0002-3344-9694}
\email{huamin@cse.ust.hk}

\author{Yukang Yan}
\authornote{This is the corresponding author.}
\affiliation{
\institution{University of Rochester}
\city{New York}
\country{United States}
}
\orcid{0000-0001-7515-3755}
\email{yukang.yan@rochester.edu}

\renewcommand{\shortauthors}{Xu and Jin, et al.}

\begin{abstract}
    360° videos enable users to freely choose their viewing paths, but blind and low vision (BLV) users are often excluded from this interactive experience. To bridge this gap, we present Branch Explorer, a system that transforms 360° videos into branching narratives—stories that dynamically unfold based on viewer choices—to support interactive viewing for BLV audiences. 
    Our formative study identified three key considerations for accessible branching narratives: providing diverse branch options, ensuring coherent story progression, and enabling immersive navigation among branches. 
    To address these needs, Branch Explorer employs a multi-modal machine learning pipeline to generate diverse narrative paths, allowing users to flexibly make choices at detected branching points and seamlessly engage with each storyline through immersive audio guidance. 
    Evaluation with 12 BLV viewers showed that Branch Explorer significantly enhanced user agency and engagement in 360° video viewing. 
    Users also developed personalized strategies for exploring 360° content. 
    We further highlight implications for supporting accessible exploration of videos and virtual environments.
\end{abstract}



\begin{teaserfigure}
  \includegraphics[width=\textwidth]{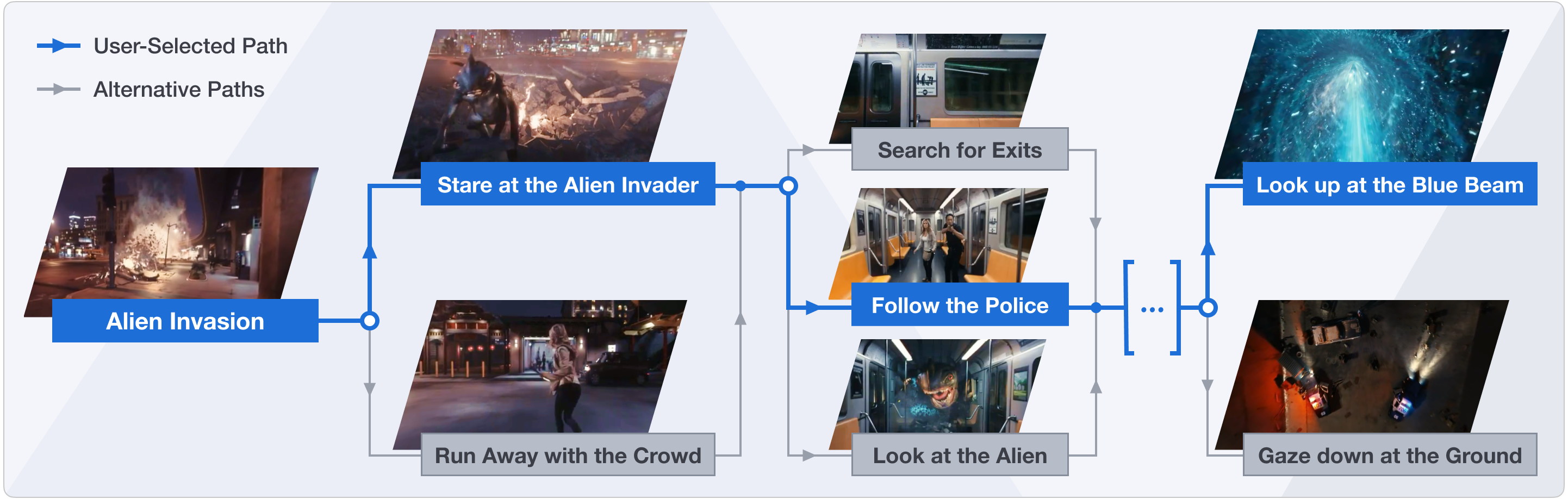}
  \caption{Branch Explorer transforms 360° videos into \textit{branching narratives}—stories that dynamically unfold based on viewer choices—to create an engaging experience for blind and low vision (BLV) users. It employs a multi-modal machine learning pipeline to generate diverse narrative paths, enabling BLV users to make choices at key branching points and explore each storyline through immersive audio guidance. The figure shows the 360° video \textit{HELP} (available at: \url{https://youtu.be/G-XZhKqQAHU}).}
  \Description{}
  \label{fig:teaser}
\end{teaserfigure}

\maketitle

\section{Introduction}

\begin{figure*}[!h]
    \centering
    \includegraphics[width=1.0\linewidth]{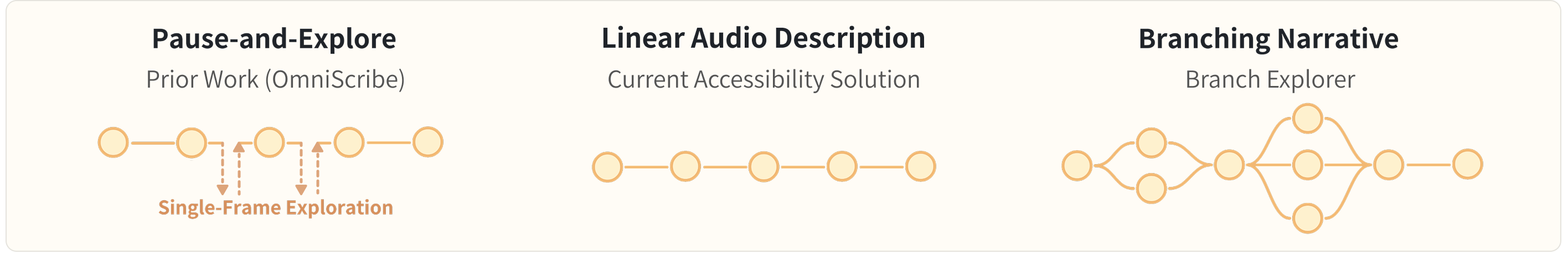}
    \caption{Comparison of the \textit{pause-and-explore} approach, \textit{linear audio descriptions}, and \textit{branching narratives} for 360° videos.}
    \label{fig:narrative_model}
\end{figure*}

360° video has revolutionized storytelling by providing users with an interactive and immersive viewing experience \cite{vskola2020virtual,linuist17outsidein,pope2017story360}. 
With a global market exceeding \$16 billion \cite{marketsizes2024vr}, 
this medium is reshaping domains from 
virtual tourism \cite{rahimizhian2020emerging,lee2024museum360}, 
online education \cite{snelson2020educational,jin2023edu360}, 
to immersive storytelling \cite{gu2016swichair,pope2017story360}. 
Unlike 2D videos that restrict viewers to a fixed perspective, 360° videos capture a panoramic field of view, 
enabling users to choose their own viewing paths \cite{amy2017uistguidance}. This interactivity encourages exploration beyond the central storyline \cite{jin2022you,wu2017dataset}, fostering a personalized viewing experience.

While 360° videos are inherently interactive, they remain largely inaccessible to blind and low vision (BLV) viewers \cite{chang2022omniscribe,jiang2023beyondAD}. 
Current accessibility solutions rely on linear audio descriptions (AD) that narrate key visual elements alongside the video in a fixed sequence \cite{wang2021Tiresias,amy2020rescribe}. 
While adequate for 2D video, this approach fails to exploit the 360° video's interactivity, reducing it to a passive experience no different from conventional media \cite{jiang2023beyondAD,chang2022omniscribe}. 
\xsc{Prior work \cite{chang2022omniscribe} took an initial step that enables BLV users to explore paused 360° frames through a pause-and-explore approach (see Figure~\ref{fig:narrative_model} left). 
However, this method interrupts the \textit{temporal continuity} of video viewing by requiring users to stop and playback, 
which breaks the natural flow of the experience. Furthermore, because interactions are limited to static frames, users cannot dynamically \textit{adjust their viewing path} during video viewing---a hallmark of 360° videos.
}

\xsc{To address this limitation, we propose using \textit{branching narratives} \cite{moser2015narrative}---a technique that dynamically adapts the storyline based on viewer choices. 
This approach is particularly well-suited to 360° video, as it provides a balance between \textit{interactivity} and \textit{narrative continuity} \cite{andrews2014visualizing,2021visualnovel,moser2015narrative}. 
It supports interactivity by allowing viewers to shape the story through multiple narrative paths, 
while maintaining a coherent experience by presenting choices at contextually appropriate moments. 
By combining these strengths, branching narratives help unlock the interactive potential of 360° video, making it a more inclusive and engaging medium for BLV audiences.}

To understand BLV viewers' needs and preferences for interacting with 360° videos via branching narratives, 
we conducted a formative study with eight BLV users who regularly watched videos. 
Participants engaged with 360° videos using a design probe that enabled branch selection via screen readers. 
Our research revealed three key considerations for accessible branching narratives: 
First, participants valued branches that vary in \textit{spatial}, \textit{semantic}, and \textit{social} aspects. 
These \textbf{diverse branches} provided users with meaningful control over the narrative. 
Second, users emphasized the importance of \textbf{coherent narration}, 
noting that abrupt pauses or disjointed content can disrupt the narrative flow and reduce the overall sense of immersion. 
Third, participants highlighted the need for \textbf{immersive navigation} across branches, 
preferring flexible interactions that do not interfere with the viewing experience.

To address these needs, we present \textit{Branch Explorer}, a system that transforms 360° videos into accessible branching narratives, enabling both \textit{interactive} and \textit{immersive} experiences for BLV users. \newline
The system incorporates three core modules: 
(1) \textbf{Branch Diversity Optimization}, 
which generates varied narrative paths, allowing users to explore the video from multiple spatial, semantic, and social perspectives; 
(2) \textbf{Coherent Narration Generation}, 
which creates consistent narrations and contextual cues to ensure a smooth narrative flow; and 
(3) \textbf{Immersive Branch Navigation}, 
which leverages subtle guidance and flexible user input to support immersive navigation among narrative paths. 
Powered by these modules, 
Branch Explorer enables users to engage with diverse storylines, 
make choices at key branching points, 
and flexibly revisit alternative branches during replay---all within an immersive experience.

To evaluate Branch Explorer, we conducted a within-subject study with 12 BLV participants, 
who compared Branch Explorer with a baseline system that simulated the state-of-the-art \textit{pause-and-explore} approach for 360° video interaction \cite{chang2022omniscribe}. 
Each participant viewed two comparable sets of videos using two systems, respectively. 
Results showed that Branch Explorer significantly enhanced users' sense of agency ($p<.01$), narrative immersion ($p<.01$), and overall engagement ($p<.05$) compared to the baseline. 
Participants reported that Branch Explorer increased the replay value of 360° videos and expressed strong willingness to use the system in the future ($\mu=6.58$ on a seven-point Likert scale). 
Users also developed \textbf{personalized strategies} for exploration, such as switching between branches to synthesize information from different perspectives or slowing down the video to track spatial changes over time. 
Based on our findings, 
we highlight implications for supporting accessible exploration of visual media in broader contexts, such as 2D videos and virtual environments.

Our contributions are threefold:

\begin{itemize}[leftmargin=*, labelindent=0pt, itemindent=0pt]
    \item We identify the needs and preferences of BLV viewers for interacting with 360° videos through branching narratives, derived from a formative study; 
    \item We present Branch Explorer, a system that incorporates three core modules---branch diversity optimization, coherent narration generation, and immersive branch navigation---to support \textit{interactive} and \textit{immersive} 360° video viewing for BLV users; 
    \item We contribute an evaluation study that demonstrates how BLV viewers interact with 360° videos using Branch Explorer and derive design implications for accessible exploration of video content and virtual environments.
\end{itemize}

\section{Related Work}
Our work extends prior research in three areas: (1) 360° video accessibility for BLV users, (2) 360° video viewing techniques, and (3) visual media accessibility for BLV users.

\subsection{360° Video Accessibility for BLV users}
For blind and low vision (BLV) users, video accessibility primarily relies on audio descriptions (AD) that narrate key visual elements alongside the video. While established guidelines exist for creating AD in traditional 2D videos \cite{li2025videoa11y,amy2020rescribe}, applying these methods directly to 360° video can restrict its interactivity, making the medium no different from conventional media \cite{360ADpractices}. 
In response, the research community has investigated new methods to make 360° videos accessible while preserving their interactive nature \cite{jiang2023beyondAD,chang2022omniscribe}.

Several studies have examined BLV viewers' preferences for describing 360° videos, revealing the need for interactive descriptions. 
For example, Fidyka et al. \cite{fidyka2018audio,fidyka2021audio,fidyka2021retelling} reported that BLV viewers wanted multiple audio descriptions for different fields of view, rather than a single focus on the main action. 
Likewise, Fleet et al. \cite{360ADpractices} found that BLV participants preferred interactive experiences over predetermined content. 
Building on this, Jiang et al. \cite{jiang2023beyondAD} conducted co-design workshops with AD creators and BLV users, highlighting the importance of subtle, lightweight guidance to help users discover relevant content within the 360° environment.

To make 360° videos accessible, prior work \cite{dang2024musical,chang2022omniscribe} has explored the use of spatial interactions. 
For example, Dang et al. \cite{dang2024musical} employed head-tracked audio descriptions to enhance BLV viewers' spatial understanding of 360° musical performances. 
Chang et al. \cite{chang2022omniscribe} developed OmniScribe, 
an AI-assisted tool that enables sighted AD creators to add spatialized object descriptions during video pauses. 
While these methods provide BLV viewers with richer spatial contexts, they also introduce challenges such as increased cognitive load, limited object discoverability, and interruptions to the narrative flow \cite{chang2022omniscribe}. Collectively, these issues can diminish the immersive experience central to 360° videos \cite{pope2017story360}.

To address these limitations, our work explores branching narratives---a lightweight and flexible approach that supports user agency in selecting viewing paths, 
while maintaining narrative coherence by introducing choices at contextually appropriate moments. 
Through a co-design process with BLV viewers, 
we identified key considerations for creating accessible branching narratives, 
and demonstrated their effectiveness in enhancing user agency and narrative immersion in 360° video experiences.

\subsection{360° Video Viewing Techniques}

360° videos enable sighted users to freely adjust their field of view within a spherical environment \cite{pope2017story360,linchi2016visualguidance}, yet viewers may miss important visual content if they look in the wrong direction \cite{amy2017uistguidance}. To address this issue, prior research has explored two main approaches: (1) interaction techniques for guiding viewer attention, and (2) automated generation of normal-view videos. 

\subsubsection{\textbf{Attention Guidance Technique}}
To direct user attention in 360° videos, prior works have explored various visual guidance techniques \cite{gu2016swichair,linchi2016visualguidance,amy2017uistguidance,linuist17outsidein,liuuist19texture,liu2023radarvr}. 
For instance, Pavel et al. \cite{amy2017uistguidance} reoriented 360° scenes to place important content within the viewer's field of view. 
Liu et al. \cite{liuuist19texture} delayed playback until the viewer looked in a salient direction by synthesizing video textures. 
Outside-In \cite{linuist17outsidein} visualized off-screen regions using picture-in-picture previews, enabling simultaneous access to multiple perspectives. 
However, existing methods primarily focus on visual consumption, offering limited support for BLV users. 
To address this gap, our work presents a non-visual attention guidance technique grounded in branching narratives. 
\hl{This approach builds on the strengths of prior visual strategies---the \textit{discoverability} afforded by multi-region previews \cite{linuist17outsidein,liu2023radarvr} and the \textit{immersion} enabled by automated view control \cite{gu2016swichair,liuuist19texture}. It further extends these paradigms by algorithmically optimizing branch diversity and timing to balance discoverability and immersion, thereby enabling both interactive and immersive 360° video experiences for BLV users.}

\subsubsection{\textbf{Normal-View Video Generation}}
Another line of work focuses on creating normal-view videos from 360° footage \cite{su2016pano2vid,shoot360,truong2018extracting,ToGnormalview,tog_360,transition360,cvprpano2d}. 
Pano2Vid \cite{su2016pano2vid} learns composition patterns from 2D video datasets to identify salient viewing directions over time. 
Shoot360 \cite{shoot360} creates normal-view videos from city panoramas using cinematic composition rules. 
Truong et al. \cite{truong2018extracting} proposed a set of user-elicited heuristics for extracting normal-view shots from 360° event videos. 
These methods are designed primarily for sighted audiences, prioritizing visual aesthetics \cite{truong2018extracting} and cinematic presentation \cite{shoot360}, without accounting for how BLV users engage with video content through audio descriptions. Our findings indicate that BLV users prefer access to multiple viewpoints that vary in \textit{spatial}, \textit{semantic}, and \textit{social} dimensions. To support these needs, we introduce an optimization-based pipeline that generates branching narratives tailored to BLV viewers' preferences.

\subsection{Visual Media Accessibility for BLV Users}
The research community has consistently focused on improving the accessibility of visual media, 
including images \cite{xu2024memoryreviver,imageExplorer2021,imageAssist}, 
videos \cite{ning2024spica,xu2025danmua11y}, 
digital games \cite{navstick,Surveyor}, 
and virtual environments \cite{Canetroller,siu2020virtual,seeingVR}. 
In this section, we review two types of media closely related to 360° content: videos and virtual environments.

\subsubsection{\textbf{Standard 2D Videos}}
To improve video accessibility, prior works have explored manual \cite{killough2023exploring}, semi-automatic \cite{liu2022crossa11y,amy2020rescribe}, 
and fully automatic methods \cite{li2025videoa11y,wang2021Tiresias} for generating audio descriptions. 
For example, VideoA11y combines multi-modal large language models and accessibility guidelines to generate descriptions tailored for BLV audiences, achieving quality comparable to trained human describers \cite{li2025videoa11y}. 
Recognizing that one-size-fits-all descriptions often fail to meet diverse user needs \cite{asset2024customAD,jiangchi24context}, 
recent research has turned toward interactive video exploration.
These solutions include providing descriptions at varying levels of granularity \cite{shortscribe}, 
supporting description customization \cite{asset2024customAD}, 
and enabling frame-level exploration \cite{ning2024spica}. 
\xsc{
However, these methods primarily require pausing the video, which interrupts the viewing experience. 
Moreover, while 2D videos have a fixed viewport, 360° videos support multiple viewing paths, presenting challenges in identifying and presenting meaningful branches. 
To address these issues, our work explores how to \textit{generate} and \textit{present} such branches based on user needs, enabling an interactive and immersive experience for BLV audiences.
}

\subsubsection{\textbf{Virtual Environments}}
Prior works have explored how to support BLV users in navigating virtual environments---including \xsc{desktop 3D video games} \cite{navstick,Surveyor} and virtual reality \cite{Canetroller,seeingVR,soundshift}. In games, accessibility tools help players scan nearby items \cite{navstick} and highlight unexplored areas \cite{Surveyor}. In virtual reality, exploration has been facilitated through multi-sensory cues such as haptic feedback \cite{Canetroller,siu2020virtual}, visual augmentations \cite{seeingVR}, auditory enhancements \cite{soundshift,may2020spotlights}, and multimodal interactions \cite{lecuyer2003homere}. While these approaches primarily focus on supporting spatial understanding, 360° videos also serve a narrative purpose---using temporally continuous footage to convey coherent stories \cite{pope2017story360}.  Our work explores how BLV users engage with these dynamic environments, revealing their unique behaviors such as slowing down videos to track spatial layout changes over time. These insights inform the design of accessible exploration methods for temporally evolving virtual environments.
\section{Formative Study}
We conducted a formative study with eight BLV viewers to understand their needs and preferences for interacting with 360° videos via branching narratives. 
Through co-watching exercises and follow-up interviews, we addressed the following questions:

\begin{enumerate}[leftmargin=*, labelindent=0pt, itemindent=0pt, label=(Q\arabic*)]
\item \textbf{Content}: \textbf{What} types of content do BLV viewers prefer in branching narratives for 360° videos? 
\item \textbf{Timing}: \textbf{When} do BLV viewers want to make branch selections during 360° video viewing? 
\item \textbf{Interaction}: \textbf{How} do BLV viewers expect to interact with branching narratives to maintain an immersive experience? 
 \end{enumerate}
\subsection{Methods}

\subsubsection{\textbf{Participants}}
We recruited eight BLV viewers\footnote{Demographic details are available in the supplementary materials.} (P1-P8; four male, four female) who regularly consumed 2D videos and had interest in 360° videos. 
Participants were recruited from an online support community, with ages ranging from 23 to 39 (mean = 30.1, SD = 5.0). 
Five participants were totally blind and three were legally blind with light perception. 
Regarding 360° video familiarity, three participants had prior viewing experience, two had only heard of them, and three had no prior knowledge. 
Additionally, six participants had previously consumed interactive stories, including interactive films (e.g., Black Mirror: Bandersnatch\footnote{\url{https://en.wikipedia.org/wiki/Black\_Mirror:\_Bandersnatch/}}) or text-based adventure games (e.g., Lifeline\footnote{\url{https://www.3minute.games/}} and TextAdventures\footnote{\url{http://textadventures.co.uk/}}). 
All participants were native Mandarin speakers.

\subsubsection{\textbf{Design Probe}}\label{sec:design_probe}
\begin{figure}[!h]
    \centering
    \includegraphics[width=1.0\linewidth]{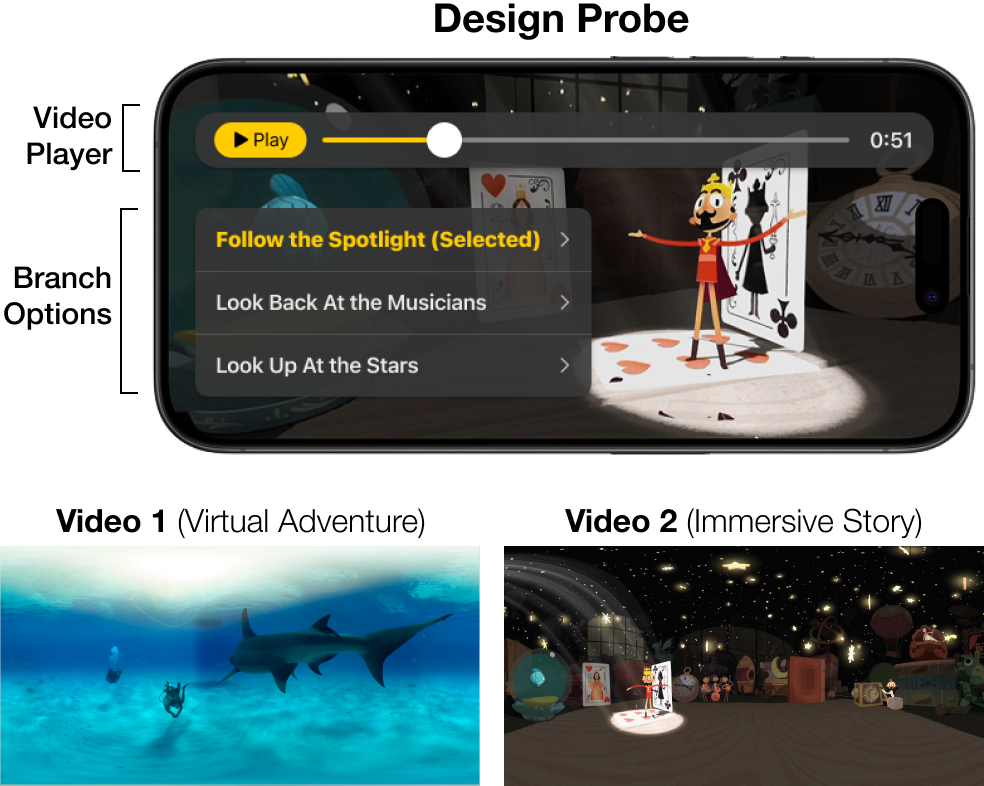}
    \caption{The design probe includes a branch option menu, allowing viewers to make choices via screen readers. Below are the two videos used in the formative study.}
    \label{fig:design_probe}
\end{figure}

We developed a design probe that enables participants to consume 360° videos via branching narratives. 
The probe is modeled after BLV users' current practices of consuming branching stories in mobile applications (e.g., Lifeline\footnotemark[2]). 
As shown in Figure~\ref{fig:design_probe}, it comprises two components: a video player and a branch option menu. 
At each branching point, the video automatically pauses, allowing viewers to make choices via screen readers \xsc{
(i.e., by touching to access the options and double-tapping to select one).} 
The video then continues with the selected branch.

\begin{figure*}[!htb]
    \centering
    \includegraphics[width=1.0\linewidth]{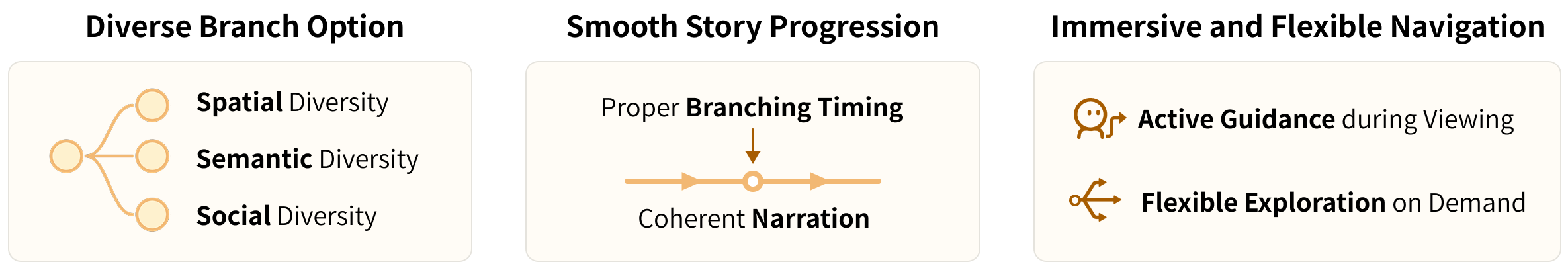}
    \caption{Three key design considerations for accessible branching narratives in 360° videos: (1) Branching options should incorporate \textit{semantic}, \textit{spatial}, and \textit{social} diversity to enhance user agency and engagement; (2) Smooth story progression is essential to avoid disruptive experiences, with \textit{choice frequency} and \textit{narrative context} being key factors in sustaining viewer immersion; and (3) Branch navigation should be immersive and flexible, \textit{adapting to different user contexts}.}
    \label{fig:formative_findings}
\end{figure*}

\xsc{The probe contains two videos that represent commonly viewed 360° video genres \cite{elmezeny2018immersive, steinhaeusser2022joyful}: 
a \textit{virtual adventure} and an \textit{immersive story} (see Figure \ref{fig:design_probe}). }
Each video is about two minutes long. 
A professional audio description creator manually crafted branching narratives for each video. 
By repeatedly reviewing the videos, the creator identified potential viewing paths and marked them as narrative branches. 
Each branch was described according to established audio description guidelines \cite{amy2020rescribe} and narrated from a second-person perspective (e.g., ``\textit{You are standing on...}'') to enhance immersion \cite{jiang2023beyondAD}. 
\hl{Prior work \cite{moser2015narrative} highlights that branching frequency is crucial to user experience. To identify a proper branching interval, each video offers three options---short (15 seconds), medium (30 seconds), and long (45 seconds)---which users select from a menu before playback.} 
The final narratives were synthesized into mono audio using the Volcengine Text-to-Speech API\footnote{\url{https://www.volcengine.com/product/tts}}.

\subsubsection{\textbf{Procedure}}
Prior to the study, participants were introduced to 360° video. 
Once they understood its panoramic nature, they proceeded with the main study, which involved two successive phases: a 20-minute co-watching exercise followed by a 40-minute interview, with a five-minute break in between.

\textbf{Phase 1: Co-watching Exercise.} 
Participants first received a brief tutorial on the design probe and then used it to view the two 360° videos in randomized order. 
They were encouraged to re-watch each video, explore different narrative branches or branching intervals, and think aloud about any confusion or suggestions.

\textbf{Phase 2: Semi-structured Interview.} 
After the co-watching session, participants first shared their experiences with the design probe. 
We then investigated their preferences regarding \textit{content}, \textit{timing}, and \textit{interaction} in branching narratives for 360° videos. 
To facilitate discussion, participants were given a list of potential topics (e.g., when branches should appear, what each branch should include, and how they prefer to select branches). 
To elicit details, participants were encouraged to support their opinions with specific examples. 
The study was conducted one-on-one and in person, and participants received the equivalent of 12 USD in local currency as compensation for their time.

\subsubsection{\textbf{Analysis}}
We recorded the study audio, transcribed it, and analyzed the data using an open-coding approach to address the three questions (Q1-Q3). 
Two researchers independently reviewed the transcripts, iteratively developing and refining the codes until they reached a consensus. 
These codes were then organized into clusters representing key themes that emerged from the data. 
After a thorough discussion of all potential themes, the final themes were reported as the study findings.

\begin{figure*}[!th]
    \centering
    \includegraphics[width=1.0\linewidth]{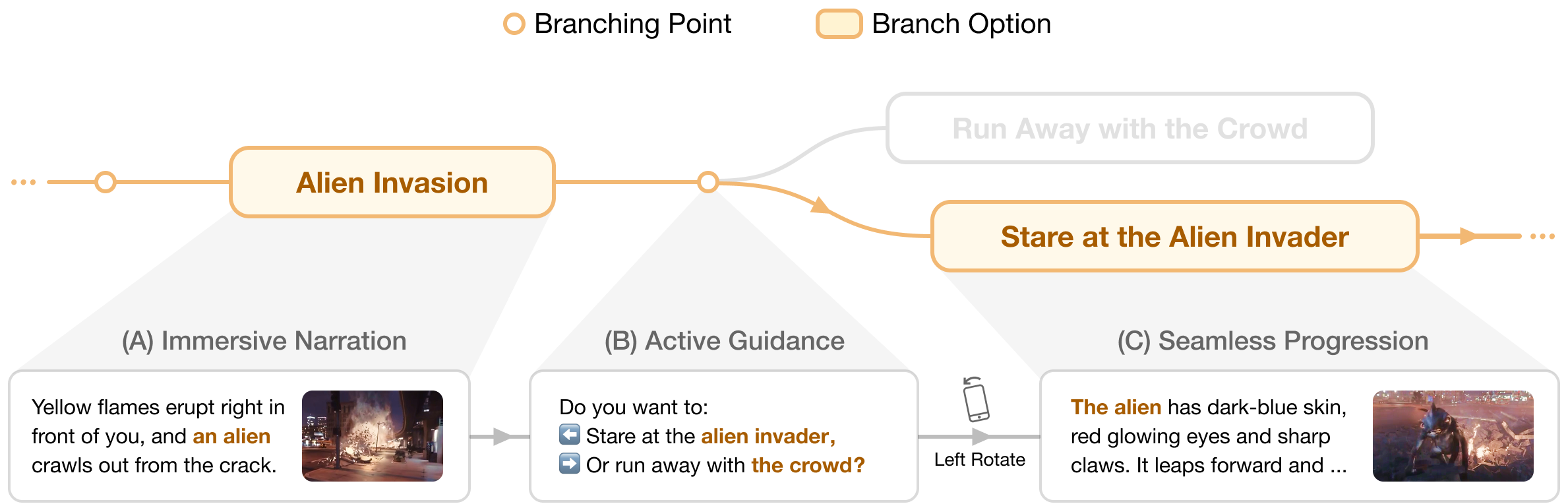}
    \caption{An example walk-through of Branch Explorer: 
    (A) The system narrates key visual elements in the second person (e.g., ``\textit{Flames erupt right in front of you...}'') to immerse users in the story. 
    (B) At branching points, users are notified through a phone vibration. 
    They can \textit{shake the phone} to reveal branch options, 
    and \textit{rotate the device} to make a choice. 
    (C) Once a branch is selected, the narration continues seamlessly along the chosen path, ensuring a smooth viewing experience.}
    \label{fig:4_1_walkthrough}
\end{figure*}

\subsection{Findings}
All participants expressed strong enthusiasm for engaging with 360° videos through branching narratives. 
They described the experience as 
``\textit{directly influencing the story}'' (P3), 
``\textit{having greater control over the narrative}'' (P8), 
and ``\textit{feeling more immersed in the virtual world}'' (P6). 
In the following, we report participants' needs and preferences in three areas: \textit{content}, \textit{timing}, and \textit{interaction}.

\subsubsection{{\textbf{Content: Diverse Branch Options}}}
Participants unanimously highlighted the importance of diverse branch options in enhancing their sense of agency. 
As P5 noted, ``\textit{When the choices are really different, I feel in control of the story.}'' 
From the interviews, we identified three key aspects of diversity:

\textbf{Spatial Diversity: Explore Different Angles.} 
Given the panoramic nature of 360° videos, all participants wanted to explore content from varied spatial perspectives. 
They appreciated options such as ``\textit{To look up at the stars or to look down at the ground}'' (P7). 
Such spatial diversity could enhance their sense of agency and improve their ``\textit{awareness of the environment}'' (P2).

\textbf{Semantic Diversity: Avoid Repetitive Content.} 
Six participants noted that when choices were identical in content but varied only by location, this led to frustration or dissatisfaction. 
For instance, P6 remarked, ``\textit{If both branches are about sharks, just in different regions, what's the point of choosing? Similar options will just bore me.}'' 
This feedback highlights the importance of providing semantically varied choices to sustain user engagement.

\textbf{Social Diversity: Reflect Public Interests.} 
Five participants expressed a desire for branches to reflect the attention of other viewers. 
For example, P1 expressed an interest in knowing, ``\textit{What are other people looking at? Are they all looking in the same direction, or are they attracted to different parts?}'' 
This awareness of public attention can ease their ``\textit{fear of missing out on interesting areas}'' (P5) and foster a sense of ``\textit{connection with others}'' (P8). Additionally, P3 expressed curiosity about little-viewed paths, noting that ``\textit{they might offer unexpected surprises}'' (P3).

\subsubsection{{\textbf{Timing: Smooth Story Progression}}}\label{sec:formative_timing}
Participants noted that smooth story progression is crucial for maintaining immersion. We identified two key considerations to achieve this:

\textbf{Proper Branching Timing.} 
Participants emphasized the importance of branching timing. 
Seven noted that branching points should \textbf{avoid overlap with speech} or important audio. 
Five preferred making choices \textbf{during scene transitions}, 
noting that ``\textit{I want to check what's new when the scene changes}'' (P5). 
Participants also highlighted the need to \textbf{avoid frequent choices}. 
\hl{When testing branching intervals of 15, 30, and 45 seconds (Section \ref{sec:design_probe}), four participants found the 15-second interval distracting, while all participants preferred the 30-second option. Based on this feedback, we set 30 seconds as the system's minimum branching interval.}

\textbf{Coherent Narration.} 
Participants highlighted the need for logical and seamless narration throughout the experience. This included ``\textit{ensuring coherent narration after choices}'' (six mentions) and ``\textit{maintaining consistent narration style and tone}'' (four mentions). Additionally, \hl{five participants reported feeling disoriented after navigating the timeline to revisit scenes or explore alternate branches.}  As P8 noted, ``\textit{After rewinding, I wonder where I am in the story.}'' 
This suggests the need for contextual cues (e.g., ``\textit{Previously...}'' recaps) to help viewers stay oriented within the narrative.

\subsubsection{{\textbf{Interaction: Immersive Navigation}}}\label{sec:formative_interaction}
Participants noted that lightweight interaction is crucial for maintaining immersion. However, current screen-reader-based solution 
``\textit{included extra steps that disrupted the immersive experience}'' (P4), 
``\textit{directly paused the video instead of letting me decide}'' (P3), 
and ``\textit{used robotic voices that didn't match the tone of the main narration}'' (P7). 
These issues prevented participants from fully engaging with the video. To address these challenges, participants shared two common expectations:

\textbf{Active Guidance during Viewing.} 
Six participants emphasized the importance of guidance during viewing because they ``\textit{often don't know when and what to explore}'' (P2). They also noted that the guidance should feel like a suggestion rather than ``\textit{taking full control}'' (P8). 
Specifically, they outlined three key expectations for effective guidance: 
(1) to signal available options subtly to avoid disruption (six \hl{mentions}), 
(2) to enable easy branch selection for smooth viewing (six \hl{mentions}), and 
(3) to maintain the same voice tone as the main narration to preserve immersion (four \hl{mentions}).

\textbf{Flexible Exploration on Demand.} Five participants expressed the need for flexible exploration during replay. 
As P8 explained, ``\textit{It's like watching movies---after an immersive experience, I want to explore more details to get a complete understanding.}'' 
Specifically, participants mentioned the importance of being able to ``\textit{access additional branches}'' (six mentions), ``\textit{explore the spatial layouts}'' (four mentions), and ``\textit{navigate among branching points}'' (three mentions). 

\subsubsection{{\textbf{Summary}}}
\xsc{Our formative study identified three key considerations for creating accessible branching narratives in 360° videos (see Figure~\ref{fig:formative_findings}): 
(1) \textbf{Content}: Branch options should incorporate \textit{semantic}, \textit{spatial}, and \textit{social} diversity to enhance user agency and engagement; 
(2) \textbf{Timing}: Smooth story progression is essential to avoid disruptive experiences, with \textit{choice frequency} and \textit{narrative context} being key factors in sustaining viewer immersion; 
and (3) \textbf{Interaction}: Branch navigation should be immersive and flexible, adapting to different \textit{user contexts}. 
These aspects are crucial to make 360° videos both interactive and engaging for BLV viewers. 
Building on these insights, we designed Branch Explorer.}

\begin{figure*}[!th]
    \centering
    \includegraphics[width=1.0\linewidth]{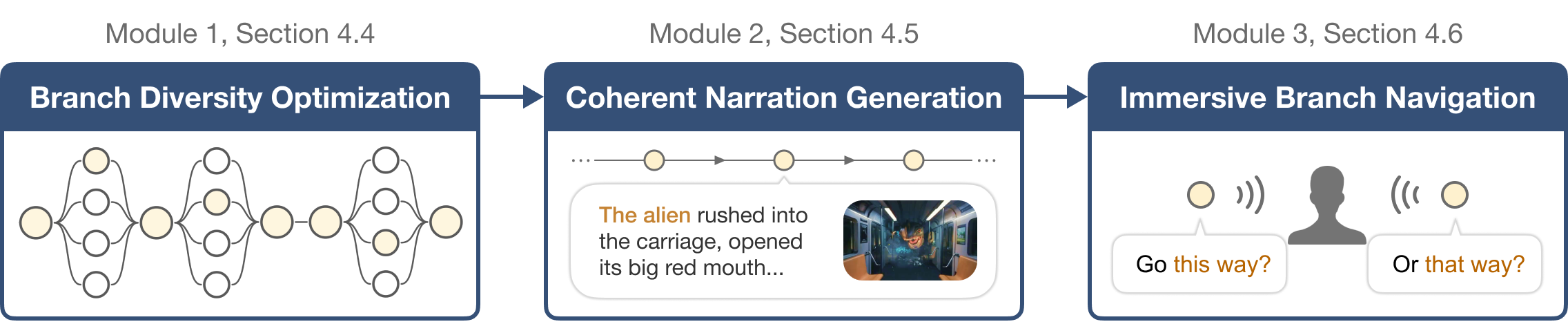}
    \caption{The pipeline of Branch Explorer. The first module employs an optimization approach to curate branches with high spatial, semantic, and social diversity. The second module generates coherent narrations and contextual cues to create a smooth narrative flow. The third module enables flexible navigation between branches, ensuring an immersive viewing experience.
}
    \label{fig:4_2_pipeline}
\end{figure*}

\section{Branch Explorer}

We present Branch Explorer, a system that transforms 360° videos into branching narratives to support interactive and immersive viewing for BLV audiences. 
It incorporates three core modules: 

\begin{enumerate}[leftmargin=*, labelindent=0pt, itemindent=0pt]
\item \xsc{\textbf{Branch Diversity Optimization}, which generates narrative branches that vary in \textit{spatial}, \textit{semantic}, and \textit{social} aspects, enabling BLV users to make meaningful choices; }
\item \xsc{\textbf{Coherent Narration Generation}, which creates coherent narrations and succinct contextual cues (e.g., plot recaps) to keep BLV users immersed in the story; and}
\item \xsc{\textbf{Immersive Branch Navigation}, which leverages subtle guidance and flexible user input to facilitate seamless navigation among narrative paths.}
\end{enumerate}

In the following sections, we first provide a user walk-through in Section~\ref{sec:system_walkthrough}. Next, we describe the system pipeline in Sections~\ref{sec:pipeline_overview}-\ref{sec:feature3}. Finally, we report the implementation details in Section~\ref{implementation Details}.

\subsection{System Walk-through}\label{sec:system_walkthrough}
\xsc{To demonstrate Branch Explorer, we follow Nick as he watches a 360° video about an alien invasion (see Figure~\ref{fig:4_1_walkthrough}). Branch Explorer enables him to interact with the story using two simple gestures: (1) shake to choose between two suggested narrative paths, and (2) swipe to browse additional options.}

\textbf{Shake-to-Choose.} 
\xsc{The video begins with a second-person narration---``\textit{You are standing in the sky, looking down at the city nightscape…}''---drawing Nick into the scene. 
When the first branching point arrives, his phone gently vibrates. 
Nick shakes the phone, triggering two options to play sequentially: ``\textit{[left ear] Stare at the invader, [right ear] or run away with the crowd?}'' 
He rotates the phone toward his left to select ``\textit{Stare at the invader}'', 
and the story proceeds along that path. 
If Nick had not responded, Branch Explorer would have defaulted to the most commonly viewed branch.}

\textbf{Swipe-to-Browse.} 
\xsc{At the next branching point, Nick is not interested in the two suggested paths---``\textit{Look at the alien}'' or ``\textit{Follow the police}''. 
He swipes to browse more options: 
``\textit{[Branch 3 of 4] Look at the poster; [Branch 4 of 4] Search for exits.}'' 
Intrigued by the idea of escaping, he double-taps to select ``\textit{Search for exits}''.
The story resumes with a quick recap---``\textit{[Previously] Ground Explosion; [Now] In the subway}''---helping Nick stay oriented without losing context.}

\subsection{Pipeline Overview}\label{sec:pipeline_overview}

As shown in Figure~\ref{fig:4_2_pipeline}, Branch Explorer incorporates three core modules: 
(1) a branch optimization module that maximizes spatial, semantic, and social diversity; 
(2) a narration generation module that produces coherent descriptions and contextual cues; and 
(3) a branch navigation module that enables flexible exploration in an immersive experience. 
The following sections first describe the method for generating branches from 360° videos, then provide details for the three core modules.

\subsection{Branch Structure Construction}

A branching narrative structure consists of two core elements (see Figure~\ref{fig:4_1_walkthrough}): (1) \textit{branching points}, where users make choices, and (2) \textit{branch options}, which allow users to select different paths. Branch Explorer constructs this structure by first identifying branching points and then generating corresponding branch options.

\subsubsection{{\textbf{Branching Point Detection}}}
Branch Explorer identifies appropriate moments for branching points based on three criteria derived from our formative study (Section~\ref{sec:formative_timing}):

\textbf{Minimize interruptions to original audio.} 
To preserve the original audio's flow, branching points are not placed during speech or loud music. 
Speech is detected using time-aligned transcripts from the Volcengine Auto-Subtitle API\footnote{\url{https://www.volcengine.com/}}. 
Loud music is identified by applying a one-second sliding window to detect regions where the root-mean-square volume exceeds a threshold of 0.8 \cite{xu2025danmua11y}. 
These regions are marked as unsuitable for branching points.

\textbf{Align with scene transitions.} 
BLV users prefer to make decisions at points where the visual scene changes. 
To accommodate this preference, 
the system segments the video into scenes using SceneDetect\footnote{\url{https://www.scenedetect.com/}}, 
a content-aware shot detection tool that compares adjacent frames in the HSV color space. Detected scene boundaries serve as initial candidates for branching points.

\textbf{Avoid frequent interruptions.} 
In our formative study, a 30-second interval between branching points was well-received by all participants. 
To enforce this, \xsc{the system iterates through branching point candidates sequentially, merging any subsequent candidates into earlier ones if they are no more than 30 seconds apart.} The resulting branch points are then used to generate branch options.

\begin{figure}[!t]
    \centering
    \includegraphics[width=1.0\linewidth]{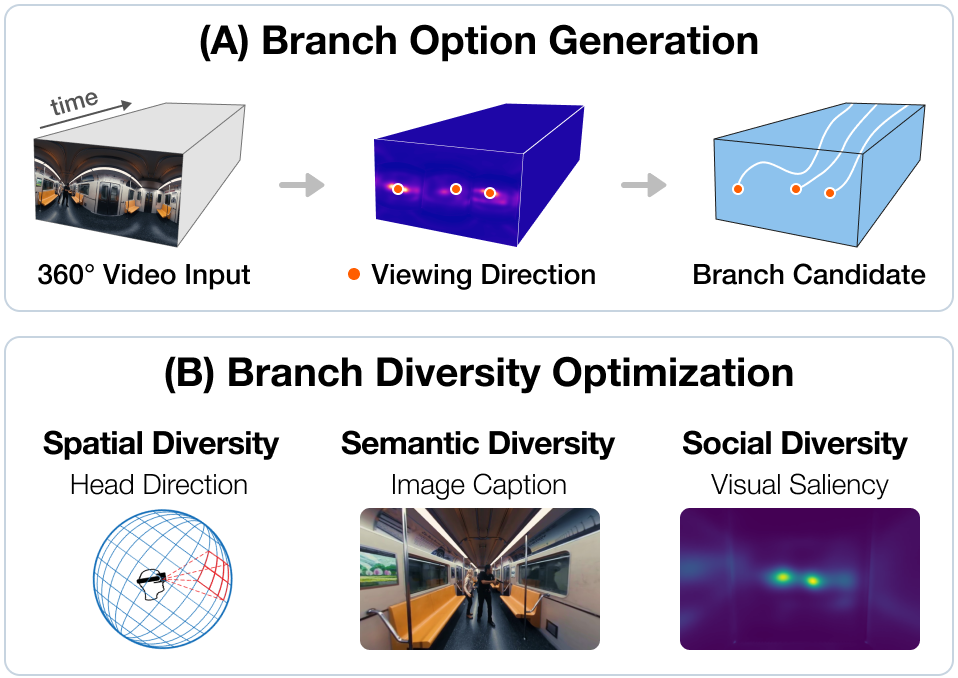}
    \caption{(A) Branch Explorer generates branches by linking frame-level viewing directions into paths. (B) It then curates branches to maximize \textit{spatial}, \textit{semantic}, and \textit{social} diversity.}
    \label{fig:diversity_optimization}
\end{figure}

\subsubsection{{\textbf{Branch Option Generation}}}
A branch consists of a series of viewing directions between two branching points. To generate branches, the system first predicts frame-level viewing directions and then links these directions into continuous viewing paths.

\textbf{Viewing Direction Prediction.} 
To predict viewing directions, we employ ATSal \cite{dahou2021atsal}, a 360° video saliency prediction model that generates frame-level saliency maps. From each map, we detect the centroids of salient regions, treating each centroid as a candidate viewing direction. To reduce redundancy, we apply agglomerative clustering with centroid linkage \cite{murtagh2012algorithms}, merging viewing directions that differ by no more than 30°. The final cluster centers represent the potential viewing directions for that frame.

\textbf{Branch Option Generation.}
The system generates branches between two branching points by linking viewing directions across frames. These directions are connected by minimizing angular differences between adjacent frames. To ensure coverage of diverse viewing paths, the process begins from the frame with the most viewing directions and propagates paths both forward and backward until reaching a branching point. \hl{A five-frame moving average filter is then applied to reduce noise in viewing paths.}

\textbf{Viewport Calculation.} The viewport for each path is defined as a 120° horizontal × 90° vertical field of view \cite{li2019very}, centered on the predicted viewing direction. The visual content within this viewport is then used for branch diversity optimization.

\subsection{Module 1: Branch Diversity Optimization}\label{sec:feature1}
BLV viewers prefer branch options that varied in spatial, semantic, and social aspects. In response, Branch Explorer employs an optimization-based approach to maximize branch diversity.

\subsubsection{{\textbf{Diversity Metrics}}}
At a given frame, the diversity between two branches is evaluated using three metrics (see Figure~\ref{fig:diversity_optimization} (B)).

\textbf{Spatial Diversity} measures the angular distance between viewing directions. Given two direction vectors, we calculate their cosine similarity $s$ and normalize it to a 0-1 scale as the spatial diversity score: $D_{\text{spa}}=0.5 \times (1-s)$. A larger score indicates higher diversity.

\textbf{Semantic Diversity} measures the semantic difference between the visual content within viewports. We quantify this using the semantic similarity between image captions. Specifically, we employ BLIP-2 \cite{li2023blip2} to caption the visual content in the 120°×90° viewport \cite{li2019very}, encode these captions into vectors using Universal Sentence Encoder \cite{cer2018universal}, and compute the cosine similarity $s'$ between the encoded vectors. The semantic diversity score is defined as: $D_{\text{sem}}=1-s'$, where a lower similarity results in a higher score. 

\textbf{Social Diversity} prioritizes branches that reflect common interests among viewers. To estimate how much a viewing direction attracts social interest, we aggregate the visual saliency values predicted by ATSal \cite{dahou2021atsal} within the 120°×90° viewport. We then apply min-max normalization across branches to produce a social diversity score $D_{\text{soc}}$ for each branch. \xsc{In the future, this score could be more accurately derived using real-user viewing data \cite{jin2022you, heatmap}.}

\textbf{Overall Diversity Score} $\textbf{\textit{D}}$ is computed as the weighted sum of three diversity metrics: \(D=w_{\text{spa}}D_{\text{spa}} + w_{\text{sem}}D_{\text{sem}} + w_{\text{soc}}D_{\text{soc}}\). For a set of branches, the semantic and spatial diversity scores are calculated as the average of pairwise scores across all branch pairs. The social diversity score is computed as the average score across all branches. All metrics are averaged over video frames and scaled between 0 and 1. We set $w_{\text{spa}} = w_{\text{sem}} = w_{\text{soc}} = \frac{1}{3}$ to assign equal weight to each metric. These weights can be further personalized based on user preferences, which we discuss in Section~\ref{sec:personalization}.

\subsubsection{{\textbf{Diversity Optimization}}}

Our formative study showed that BLV viewers have varying preferences for the number of branch options: they prefer essential options during immersive viewing but desire access to more branches during replay. To accommodate these needs, the system iteratively selects branches that maximize the diversity score. It begins by selecting the branch \( B_0 \) with the highest social diversity score: 
\(B_0 = \arg\max_{B} D_{\text{soc}}(B)\). 
The system then iteratively adds a new branch \( B_k \) that maximizes the overall diversity: 
\(B_k = \arg\max_{B \notin \{B_0, \dots, B_{k-1}\}} D(B_0, \dots, B_{k-1}, B)\). 
The process stops when adding a new branch reduces the diversity score by a certain margin: 
\(D(B_0, \dots, B_k) < \lambda D(B_0, \dots, B_{k-1})\). 
We set $\lambda=0.75$ to discard branches that significantly reduce the overall diversity. The final output is a sorted list of branches, enabling the system to present the top $N$ options, where $N$ can be adjusted based on user preferences. To avoid overwhelming users, each branching point was limited to a maximum of five options \cite{hick1952rate,miller1956magical}. \xsc{This maximum may restrict coverage in scenes with dense areas of interest (e.g., people appearing in all directions); we discuss potential methods to mitigate this issue in Section~\ref{sec:pipeline_improvement}.}

In summary, this approach prioritizes branches with high diversity, gradually adds new ones when more options are needed, and filters out those too similar to existing branches---thereby balancing the need for diverse choices with the desire to avoid long pauses.

\subsection{Module 2: Coherent Narration Generation}\label{sec:feature2}
Branch Explorer generates \textit{coherent narrations} and \textit{contextual cues} to provide a seamless viewing experience for BLV viewers.

\subsubsection{{\textbf{Branch Narration Generation}}}
\hl{To ensure a coherent narrative flow, the system adopts a \textbf{continuation writing strategy}, which generates branch narrations by \textit{continuing from the preceding branches} rather than treating each segment in isolation. This approach enables each branch to build coherently on prior content.
For each branch, a 120°×90° viewport video clip is processed using the VideoA11y pipeline \cite{li2025videoa11y},  which identifies keyframes based on luminance peaks and uses GPT-4o to generate descriptions.} 
Specifically, the pipeline receives narrations from all preceding branches and is prompted to ``\textit{generate descriptions that are coherent with previous content}''\footnote{Prompts are available in the supplementary materials.}. 
To enhance immersion, the model is instructed to ``\textit{narrate in a second-person perspective}'' \cite{jiang2023beyondAD}. 
To ensure quality, the model is guided by 42 audio description guidelines drawn from prior work \cite{li2025videoa11y} (e.g., ``\textit{Describe essential visual content objectively.}''). 
The resulting narration is inserted into the longest non-speech segment of each branch, as detected by the Volcengine Auto-Subtitle API. The narration length is constrained by this interval, using a speech rate of approximately three Chinese words per second.

\begin{figure}[!b]
    \centering
    \includegraphics[width=1.0\linewidth]{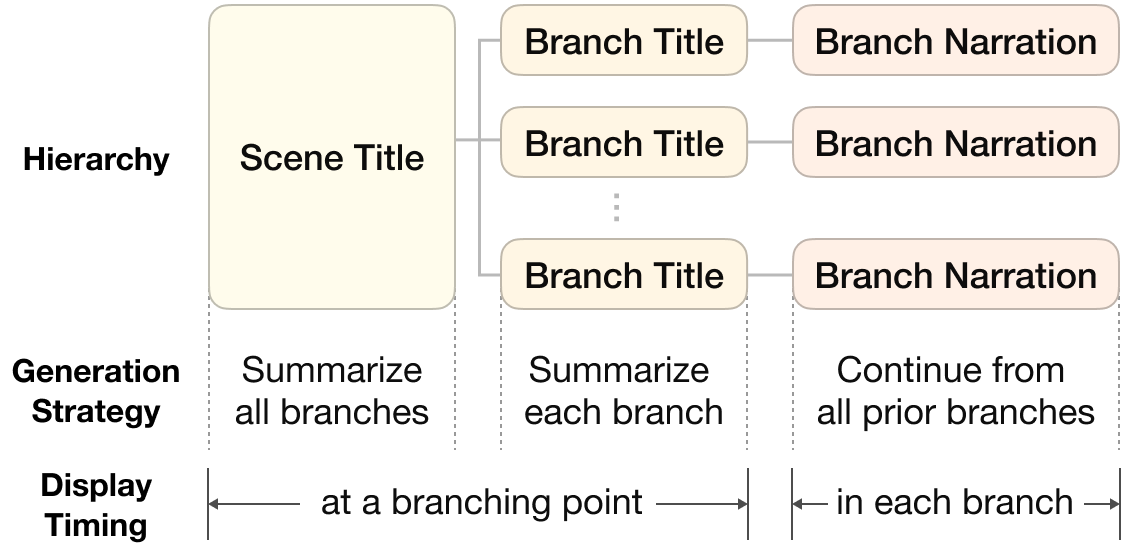}
    \caption{\hl{The system generates \textit{scene titles}, \textit{branch titles}, and \textit{branch narrations} to ensure narrative coherence. This figure illustrates their hierarchy, generation strategy, and display timing (i.e., when they are provided to users).}}
    \label{fig:feature2_coherence}
\end{figure}

\subsubsection{{\textbf{Navigation Cue Generation}}} 
\xsc{Participants in the formative study relied on \textbf{navigation cues} (i.e., branch and scene titles) to stay oriented within the storyline. 
In response, the system uses GPT-4o to generate these cues. 
At each branching point, GPT-4o receives the full set of branch narrations and is prompted to: 
``\textit{[Branch Title] Generate a summary for each branch; 
[Scene Title] Generate a summary for all the branches}''. 
Figure~\ref{fig:feature2_coherence} shows the hierarchy, generation strategy, and display timing for these descriptions.}

\subsection{Module 3: Immersive Branch Navigation}\label{sec:feature3}

\xsc{Our formative study revealed that BLV users' interaction preferences vary based on the viewing context: they prefer subtle guidance during initial viewing and seek greater flexibility during replay. 
In response, 
Branch Explorer offers \textbf{three exploration features} that range from lightweight guidance to in-depth exploration, 
enabling immersive navigation throughout the viewing experience.}

\subsubsection{{\textbf{Exploration Feature Overview}}}
\xsc{Branch Explorer enables 360° video exploration at three levels: 
(1) \textit{Branch Selection}: At branching points, the system uses phone vibrations to signal options, allowing users to select a path by simply shaking the phone. 
(2) \textit{Storyline Navigation}: Users can swipe at any time to reveal additional branches and jump between branching points. 
(3) \textit{Spatial Exploration}: During pauses, users can turn their heads to trigger object descriptions. 
Each feature uses a distinct input modality---shaking, swiping, and head turning---to ensure smooth interaction.}

\begin{figure}[!b]
    \centering
    \includegraphics[width=1.0\linewidth]{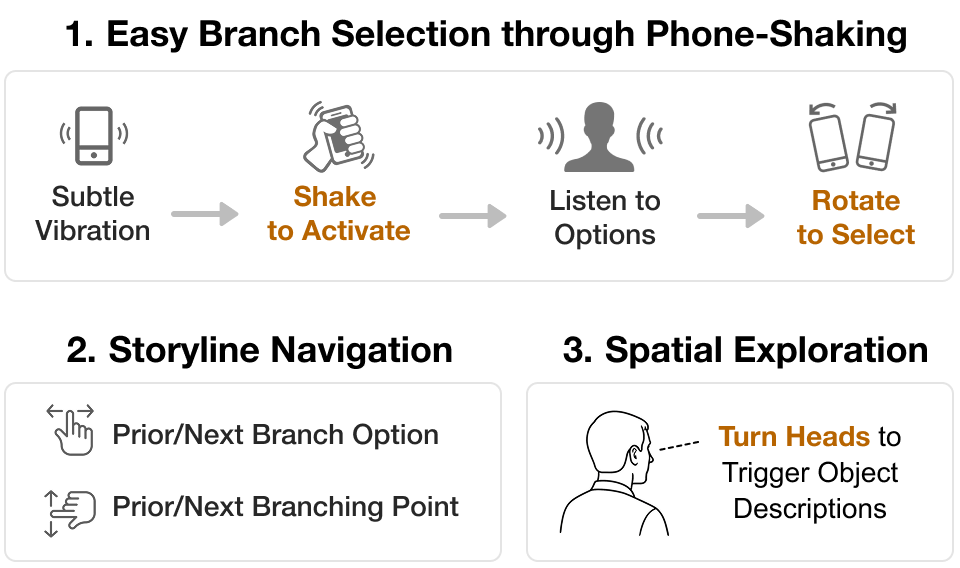}
    \caption{Branch Explorer provides three exploration features: (1) \textit{branch selection} via shaking, (2) \textit{storyline navigation} via swiping, and (3) \textit{spatial exploration} via head-turning.}
    \label{fig:feature3_interaction}
\end{figure}

\subsubsection{{\textbf{Feature 1: Easy Branch Selection}}} 
\xsc{The system offers subtle guidance for branch selection. 
At each branching point, the phone briefly vibrates for half a second. 
If users want to make a choice, they can simply \textbf{shake the phone} within five seconds. 
Otherwise, the video continues automatically along the most commonly viewed branch, 
ensuring a seamless experience. 
\hl{We selected the shaking gesture as a simple interaction method preferred by BLV users \cite{xu2025danmua11y}. 
} 
After a phone shake, the video pauses and presents two options using spatial audio---first from the left, then from the right. 
Users select a branch by \textbf{rotating the phone} in the corresponding direction, 
and the video resumes smoothly along the chosen path.
}

\subsubsection{{\textbf{Feature 2: Flexible Storyline Navigation}}} 
\xsc{
The system supports storyline navigation along two axes: 
(1) switching between narrative branches, and 
(2) jumping across branching points. 
At any time, users can \textbf{swipe} left or right with one finger to cycle through an optimized list of branches, or swipe up or down to move between branching points. 
When a swipe is detected, the system pauses the video and provides a navigation cue to indicate the user's current location in the storyline. 
Informed by prior work \cite{xu2024memoryreviver}, the system announces the current location using numbered identifiers (e.g., ``\textit{[Scene 3 of 7] In the subway; [Branch 3 of 3] Search for Exits}''). 
After users confirm their choice with a double tap, the video resumes from the beginning of the chosen branch, accompanied by a brief summary to re-establish the narrative context (e.g., ``\textit{[Previously] Ground Explosion; [Now] In the subway}''). 
}

\subsubsection{{\textbf{Feature 3: Detailed Spatial Exploration}}} 
\xsc{
To enable BLV users to explore the spatial layouts of a scene, Branch Explorer integrates a spatial exploration feature inspired by prior work \cite{chang2022omniscribe}. When the video is paused (i.e., by a double tap), users can explore the scene by simply \textbf{turning their heads}. As the head direction enters a specific viewport, the system delivers an object description. 
\hl{These descriptions are only provided during video pauses to avoid overlapping the main narration \cite{amy2020rescribe}. }
Descriptions are image captions generated using BLIP-2 \cite{li2023blip2}, 
based on a 120°×90° viewport for each branch, 
sampled at one frame per second. 
If the user’s head direction aligns with multiple overlapping viewports, the system selects the one with the smallest angular deviation to ensure the most relevant description is provided.
}

\subsubsection{{\textbf{Video Control}}} 
In addition to the three exploration features, the system supports standard video controls. Users can pause or play the video with a one-finger double tap, and fast-forward or rewind by five seconds using a two-finger swipe up or down, respectively.

\subsection{Implementation}\label{implementation Details}
\subsubsection{\textbf{System}}
The system is implemented as an iPhone application, tested on an iPhone 15 with AirPods Pro (second generation). Phone gestures and head movements are detected using the \href{https://developer.apple.com/documentation/coremotion}{CoreMotion API}. Audio is managed through AVFoundation. 
All textual content is synthesized into speech using the \href{https://www.volcengine.com/product/tts}{Volcengine Text-to-Speech API}, 
with the same human-like voice tone\footnote{The tone ID is BV411 in Volcengine Text-to-Speech API.}. 
The branch narration is rendered at a speech rate between 1.1 and 1.2, adjusted to fit non-speech intervals estimated by word count. 
\hl{The narration is presented in mono audio to minimize cognitive load from excessive spatial audio cues \cite{chang2022omniscribe, jiang2023beyondAD}.}

\subsubsection{\textbf{Pipeline}}
The video processing pipeline is implemented in Python. The video is sampled at one frame per second for branch generation and optimization. Diversity metrics are precomputed and passed into an optimization module implemented using dynamic programming. For modules utilizing GPT-4o, the temperature is set to 0.2 to encourage deterministic descriptions \cite{yang2023dawn}.

\section{Technical Evaluation}\label{sec:technical_eval}
We evaluated Branch Explorer using the 360° video dataset from \cite{wu2017dataset}, which contains eighteen videos\footnote{Video details are available in the supplementary materials.} across five genres: film, documentary, sport, talk show, and performance. The videos range from two to ten minutes in length and include both static scenes and complex motion (e.g., skiing). \hl{Using the pipeline output, we evaluated our system in three aspects: \textit{branching timing}, \textit{branch diversity}, and \textit{description accuracy}.}

\subsection{\hl{Branching Timing and Diversity}}
\hl{Branching timing and branch diversity were evaluated using annotations provided by professional audio describers.}

\subsubsection{\textbf{\hl{Branching Timing.}}} 
\hl{To assess branching timing, we compared our system's branching points in four videos (V1–V4 in Figure~\ref{fig:eval_video}) with labels independently marked by two professional audio describers (Experts M and N). Branching points within five seconds of each other were considered equivalent. We used the Jaccard index to quantify similarity. The agreement rates were 57.6\% between Experts M and N, 51.5\% between the system and M, and 61.1\% between the system and N. These results suggest that the system's alignment with human experts was comparable to the inter-expert agreement. Most matched branching points occurred at major scene changes (e.g., entering a new location), while discrepancies mainly reflected individual preferences for branching frequency in complex scenes (e.g., branching once for an entire subway scene vs. branching three times for entering, being in, and leaving the subway).}

\subsubsection{\textbf{\hl{Branch Diversity.}}} 
\hl{To evaluate branch diversity, we randomly sampled twenty sets of branches from the pipeline output. 
A professional audio describer (Expert M) rated the diversity of each set using a seven-point Likert scale based on the following prompt: ``\textit{The branch options were highly diverse and covered most salient regions.}'' 
The resulting average rating was $\mu$ = 6.65 (SD = 0.57), suggesting that the system generated branches with high diversity. Expert M noted that manually creating such varied options would be tedious and praised the system for surfacing subtle details (e.g., small objects) that might otherwise be overlooked.}

\subsection{Description Accuracy}
We assessed the accuracy of three types of descriptions generated by the system: (1) branch narrations, (2) navigation cues (i.e., scene and branch titles), and (3) object descriptions.

\subsubsection{\textbf{Methods.}} 
For each description type, we randomly sampled 100 items from the pipeline output. Errors were annotated by reviewing each description alongside its corresponding video clip. A description was marked incorrect if it did not match the visual content. One researcher performed the initial labeling, and a second researcher reviewed the labels to ensure reliability. Accuracy was calculated as the percentage of correct descriptions among the 100 sampled items for each type.

\begin{table}[!b]
\caption{Accuracy rate, word count, and error count for each description type.  ($\mu$ = mean, $\sigma$ = standard deviation)}
\label{tab:accuracy_stats}
\centering
\resizebox{1.0\columnwidth}{!}{
\setlength{\tabcolsep}{1.6mm}{
\renewcommand\arraystretch{1.2}
\newcommand{\hlineblack}{\specialrule{0.1em}{0em}{0em}}
\begin{tabular}{c|c|>{\centering\arraybackslash}p{0.7cm}>{\centering\arraybackslash}p{0.7cm}|>{\centering\arraybackslash}p{0.7cm}>{\centering\arraybackslash}p{0.7cm}}
\hlineblack
\multirow{2}{*}{\textbf{Type}} & \multirow{2}{*}{\textbf{Accuracy}} & \multicolumn{2}{c|}{\textbf{Word Count}} & \multicolumn{2}{c}{\textbf{Error Count}} \\
\cline{3-4} \cline{5-6}
& & $\mu$ & $\sigma$ & $\mu$ & $\sigma$ \\
\hlineblack
Branch Narration       & 78\% & 74.8 & 18.7 & 0.60 & 1.32 \\
Navigation Cue     & 96\% & 6.8  & 2.0  & 0.04 & 0.20 \\
Object Description     & 89\% & 12.6 & 3.4  & 0.11 & 0.31 \\
\hlineblack
\end{tabular}
}
}
\end{table}

\subsubsection{\textbf{Results.}} 
Table~\ref{tab:accuracy_stats} shows the accuracy rates and error counts for each description type. 
Navigation cues, object descriptions, and branch narrations achieved an accuracy of 96\%, 89\%, and 78\%, respectively. 
These results align with prior studies \cite{shortscribe,liu2024survey}, 
which suggest that longer descriptions are more prone to hallucinations.

\hl{We further analyzed branch narration accuracy by video genre. Accuracy rates were higher in static scenes (talk show: 91\%, performance: 85\%) compared to dynamic ones (film: 77\%, documentary: 74\%, sport: 71\%). 
The primary sources of error included rapid object motion, 
non-ideal composition (e.g., objects near the viewport edge), 
and inconsistent object references across branches.} 
Future work may mitigate these issues by tracking fast-moving objects to describe their movements \cite{zhang2022bytetrack}, 
incorporating composition constraints during branch generation \cite{shoot360,truong2018extracting}, 
and enhancing reference consistency through spatial-temporal video grounding \cite{liu2024single}, 
which we discuss in Section~\ref{sec:pipeline_improvement}. 
\section{User Evaluation}
We conducted a within-subject study with 12 BLV viewers to evaluate the effectiveness of Branch Explorer compared to a baseline. Specifically, we aimed to address the following research questions:

\begin{enumerate}[leftmargin=*, labelindent=0pt, itemindent=0pt, label=(RQ\arabic*)]
\item \textbf{System Usability}: How do BLV users perceive the usability of Branch Explorer when watching 360° videos?
\item \textbf{Agency and Immersion}: How does Branch Explorer affect users' sense of agency and immersion during video viewing?
\item \textbf{Video Comprehension}: How does Branch Explorer impact BLV viewers' understanding of video content?
\item \textbf{Exploration Strategy}: When and how do BLV viewers explore 360° videos using Branch Explorer?
\end{enumerate}

\subsection{Participants and Materials}
\subsubsection{\textbf{Participants}}
We recruited 12 BLV viewers\footnote{Demographic details are available in the supplementary materials.} (P9-P20; six male, six female) who regularly consumed 2D videos and had interest in 360° videos. 
They were recruited from an online support community, with ages ranging from 25 to 45 (mean = 32.3, SD = 6.0). 
Seven participants were totally blind and five were legally blind with light perception. 
Four participants had prior experience viewing 360° videos, five had only heard of them, and three had no prior knowledge. 
Ten participants had previously consumed interactive stories. 
None of the participants participated in the formative study.

\subsubsection{\textbf{Baseline}}
To evaluate the impact of branching narratives on the 360° video viewing experience, we implemented a baseline system by removing all branching-related features from Branch Explorer. 
This condition simulated the state-of-the-art pause-and-explore approach for 360° video interaction \cite{chang2022omniscribe}. 

Specifically, the baseline incorporated the following functionalities: 
(1) narration of key visual elements alongside the video, 
(2) phone vibrations to signal scene transitions (i.e., identified decision points), 
(3) double-tap gestures to play or pause the video, 
(4) head-turning during pauses for spatial exploration, and 
(5) swipe up/down gestures for timeline navigation. 
For each scene, narration was selected from the branch with the highest social diversity score, reflecting the most common viewing path.

\subsubsection{\textbf{Apparatus}}
Both systems were deployed on an iPhone 15. Participants wore AirPods Pro to receive audio during video viewing. To facilitate head movements for spatial exploration, participants were seated in a swivel chair throughout the study.

\subsubsection{\textbf{Videos}}
\begin{figure}[!h]
    \centering
    \includegraphics[width=1.0\linewidth]{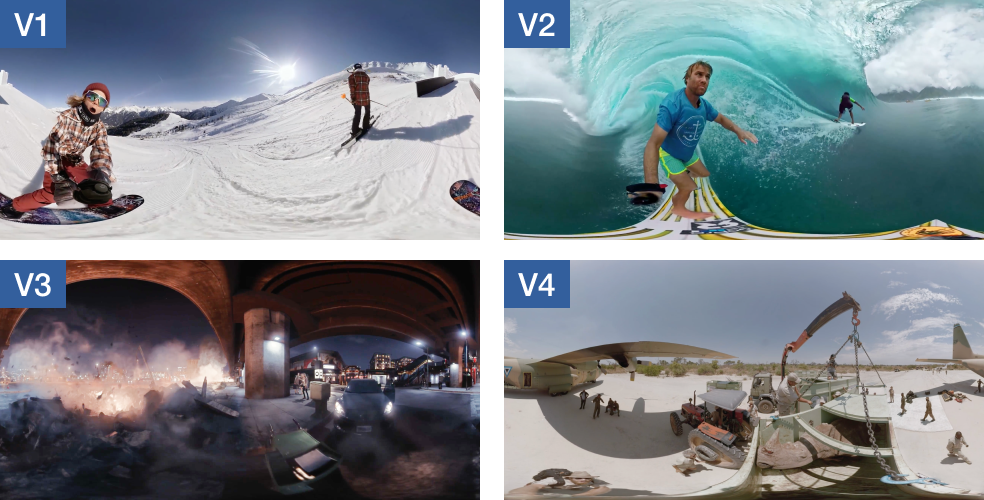}
    \caption{The four videos used in the evaluation study.}
    \label{fig:eval_video}
\end{figure}

To ensure a fair comparison, we selected four videos (V1-V4 in Figure~\ref{fig:eval_video}) from a 360° video dataset \cite{wu2017dataset}. 
These videos were divided into two groups with comparable durations and content types. 
In each group, one video featured virtual adventures, such as skiing (V1/V2, about 3.5 minutes long), 
while the other focused on immersive stories, such as films (V3/V4, about five minutes long). 
To avoid misleading users, we manually corrected errors in descriptions only when they contradicted the visual content.

\subsubsection{\textbf{Comprehension Questions}}
To evaluate BLV viewers' understanding of the main video content, we designed four comprehension questions per video. These questions focused on key actions and characters (e.g., ``\textit{The alien looks like a dinosaur.}'') and were answerable regardless of the chosen narrative path. Participants selected from three response options---``\textit{yes}'', ``\textit{no}'', or ``\textit{I'm not sure}''---with ``\textit{I'm not sure}'' responses counted as incorrect.

\subsection{Design and Procedure}
\subsubsection{\textbf{Procedure}}
Participants first provided their demographics 
and shared their prior experiences with 360° videos and interactive stories. 
They then received a 10-minute tutorial on both systems 
using the same video (V5 in the supplementary materials). 

After the tutorial, the main study began. 
Each participant used both systems to view 360° videos. 
To ensure study control, the sequence of systems and video groups was counterbalanced by creating four combinations, 
which were evenly assigned among the twelve participants. 
The videos were presented in a random order. 
Participants were instructed to watch videos as they would in daily life---pausing, fast-forwarding, rewinding, or utilizing the exploration features as needed---and to replay videos if they wished to explore further. 
They were also encouraged to verbalize their \textbf{motivations for exploration} during viewing. 
After each video, participants composed a brief \textbf{video summary} and 
then answered four \textbf{comprehension questions}.

\begin{figure*}[h]
    \centering
    \includegraphics[width=\textwidth]{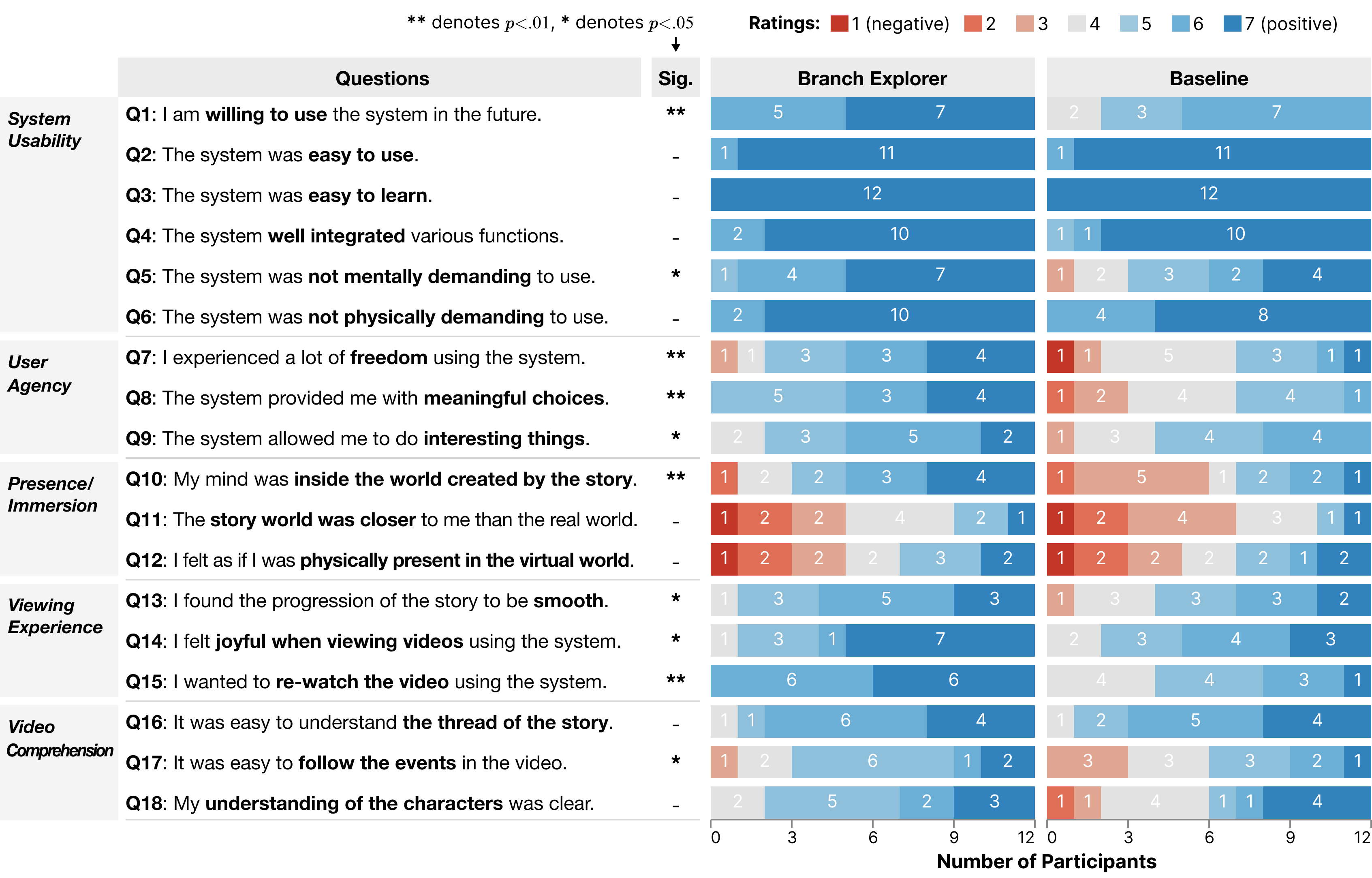}
    \caption{Distributions of participant ratings for both systems (1=strongly negative, 7=strongly positive). Asterisks denote statistical significances based on the Wilcoxon signed-rank test (Detailed results are available in the supplementary materials).}
    \label{fig:comparison_ratings}
\end{figure*}

Upon completing both systems, participants filled out a questionnaire (see Figure~\ref{fig:comparison_ratings}), followed by a semi-structured interview to gather their feedback. The study lasted about 90 minutes and was conducted one-on-one, in-person. 
Participants were compensated approximately 18 USD in local currency for their time.

\subsubsection{\textbf{Metrics}}
We used comprehension metrics and subjective ratings to assess both systems.

\textbf{Comprehension Metrics.} 
We assessed video comprehension using three metrics:
(1) the \textit{accuracy rate} on comprehension questions,
(2) the \textit{word count} of participants' video summaries, and
(3) the \textit{number of errors} identified in those summaries.

\textbf{Subjective Ratings.} 
We adapted established scales \cite{lewis2018system,ryan2006motivational,busselle2009measuring,hartmann2015spatial} to address our research questions. 
For usability, we assessed participants' willingness to use the system in the future, ease of use, and ease of learning using items from the System Usability Scale \cite{lewis2018system}. 
User agency was evaluated using the autonomy sub-scale from the Player Experience of Need Satisfaction Scale \cite{ryan2006motivational}. 
For immersion, we measured narrative and spatial presence using questions from the Narrative Engagement Scale \cite{busselle2009measuring} and the Spatial Presence Experience Scale \cite{hartmann2015spatial}, respectively. 
Video comprehension was measured using narrative understanding items from the Narrative Engagement Scale \cite{busselle2009measuring}. 
All questions used a 7-point Likert scale.

\subsubsection{\textbf{Analysis}}
We collected audio recordings, interaction logs, and questionnaire responses during the study. 
For quantitative data that satisfied normality and equal-variance assumptions, we conducted paired t-tests to assess statistical significance. 
Subjective ratings were analyzed using the Wilcoxon signed-rank test \cite{wilcoxon1970critical}. 
To evaluate the number of errors in video summaries, one researcher shuffled the content to obscure participant and system details. A second researcher, who was unaware of this information, reviewed the summaries and annotated any factual errors. The audio recordings were transcribed and categorized according to the research questions. The findings are reported based on this analysis.
\section{User Evaluation Results}
We conducted 48 trials (12 participants $\times$ 2 videos $\times$ 2 systems) in total. 
Participants re-watched videos if they wished to explore further using the system. 
On average, each video was viewed 2.20 times (SD=0.49)\footnote{Calculated by dividing the overall play time (excluding pauses) by the video duration.} 
using Branch Explorer, 
compared to 1.73 times (SD=0.33) with the baseline. 
In the following sections, we present our findings on system usability, agency and immersion, video comprehension, and exploration strategy.

\subsection{System Usability (RQ1)}
\textbf{Branch Explorer achieved high usability by seamlessly integrating three exploration features}---branch selection, storyline navigation, and spatial exploration. 
Participants described these features as ``\textit{intuitive and well organized---shake, swipe, and turn around---each for a different purpose}'' (P19). They also appreciated how the features progressed ``\textit{from simple to more advanced, supporting both smooth viewing and detailed exploration}'' (P15). Consequently, Branch Explorer was rated as highly easy to learn ($\mu = 7.00$ out of 7), easy to use ($\mu = 6.92$), and well-integrated across functions ($\mu = 6.83$), leading to a significantly higher willingness to use the system in the future ($Z = -2.89,\ p < .01$) compared to the baseline.

\textbf{Branch Explorer reduced the mental effort during exploration by making available actions discoverable.} Participants rated the system as significantly lower in mental demand compared to the baseline ($Z = -2.46,\ p < .05$), noting that ``\textit{It made me more aware of when and what to explore---The vibration tells me there are some options, so I can shake to access them.}'' (P17). In contrast, while participants appreciated the richer detail offered by spatial exploration, they often had to ``\textit{pause to see if there's something new}'' (P12) and ``\textit{turn around to search for objects}'' (P9). This led to prolonged exploration and increased resumption lag \cite{altmann2004task}---sometimes even ``\textit{forgetting the main narrative}'' (P13). These findings highlight the importance of making actions easily discoverable to reduce the cognitive load during exploration. Future systems could enhance object discovery during spatial exploration by using audio cues \cite{rothe2018guiding} or by presenting structured object lists via screen readers \cite{imageAssist}.

\begin{figure*}[h]
    \centering
    \includegraphics[width=\textwidth]{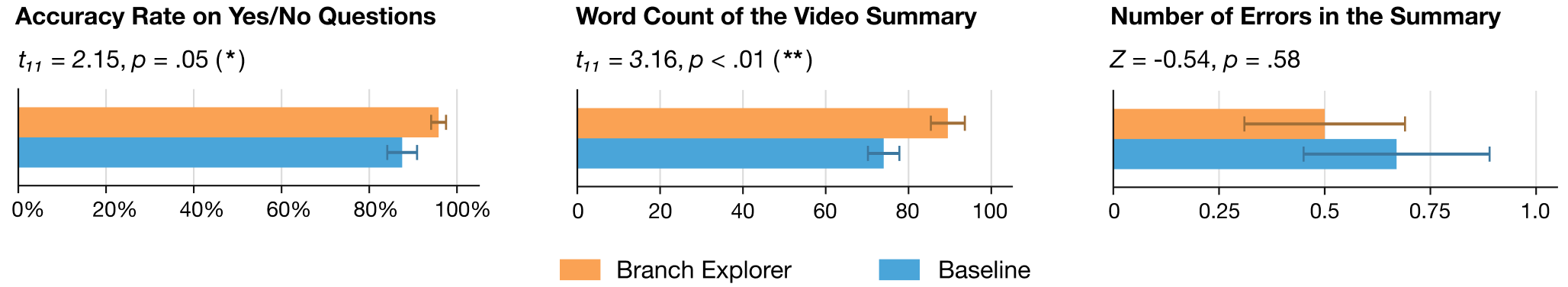}
    \caption{Video comprehension metrics from the evaluation study. These values were computed as \textbf{averages per participant for each video}. Asterisks denote statistical significances. Error bars represent standard errors.}
    \label{fig:comprehension_metrics}
\end{figure*}

\subsection{Agency and Immersion (RQ2)}
\textbf{Branch Explorer significantly improved BLV users' sense of agency during video viewing.} Participants reported experiencing a greater sense of freedom ($Z=-3.03,\ p<.01$), feeling more able to make meaningful choices ($Z=-3.11,\ p<.01$), and engaging in more interesting actions ($Z=-2.53,\ p<.05$). They attributed this to the system ``\textit{providing distinct options}'' (P17), enabling them to ``\textit{directly shape the story}'' (P18). Participants also noted that the simple interactions made them ``\textit{more willing to explore}'' (P11).

\textbf{Importantly, the system's active guidance did not take over user control.} Participants described the vibration notifications as ``\textit{a gentle reminder, not a command}'' (P9), which allowed them to ``\textit{control the viewing pace}'' (P14). However, some expressed a desire for greater freedom, such as ``\textit{making decisions at any time}'' (P15). This highlights the need for adaptive descriptions that respond to user intent in real time, which we discuss in Section~\ref{sec:dis_context}.

\textbf{Branch Explorer significantly enhanced users' sense of narrative presence.} Participants reported feeling more present in the story world compared to the baseline ($Z = -2.87,\ p < .01$), describing themselves as ``\textit{actively participating in the story}'' (P10) and ``\textit{deeply caring about how the narration continues}'' (P19). They noted that ``\textit{the seamless narrative flow made the video more immersive}'' (P14). However, this narrative immersion did not translate into stronger spatial presence: they neither perceived the story world as closer than the real world ($Z = -1.73,\ p = .08$) nor felt physically situated within the virtual environment ($Z = -0.38,\ p = .71$). Participants reported several barriers, including ``\textit{the audio effects not resembling a physical soundscape}'' (P12) and ``\textit{the spatial exploration via turning heads not mimicking how we navigate the real world}'' (P14). These findings suggest that future systems could incorporate realistic spatial audio \cite{spatial_audio} and real-world interactions such as blind photography \cite{jayant2011supporting} to enhance spatial presence.

\textbf{Consequently, participants reported greater enjoyment while viewing videos with Branch Explorer.} They rated the story progression as significantly smoother ($Z = -2.53,\ p < .05$), felt more joyful ($Z = -2.12,\ p < .05$), and expressed a stronger willingness to re-watch videos using the system ($Z = -3.02,\ p < .01$). Participants highlighted that Branch Explorer ``\textit{offered unique engagement compared to traditional interactive stories}'' (P13), as its focused on ``\textit{providing different ways for interpreting the same event}'' (P15). This design made them feel like they were ``\textit{playing a detective game---piecing together clues to form a full picture}'' (P10), which ``\textit{increased the overall engagement of video viewing}'' (P16).

\subsection{Video Comprehension (RQ3)}

\textbf{Branch Explorer improved viewers' video comprehension.} 
\xsc{Figure~\ref{fig:comprehension_metrics} shows the results on comprehension metrics. On average, participants achieved a higher accuracy rate on comprehension questions with Branch Explorer ($\mu=95.8\%,\ \sigma = 2.4\%$) compared to the baseline ($\mu=87.5\%,\ \sigma = 5.2\%$; $t_{11} = 2.15,\ p = .05$).} 
\hl{The majority of errors were due to selecting ``\textit{I'm not sure}'' (83\% in the baseline and 75\% with Branch Explorer)}, which participants attributed to ``\textit{losing context after long pauses}'' (P13) or ``\textit{unnoticed details}'' (P18). 
Participants also wrote significantly longer video summaries with Branch Explorer ($t_{11} = 3.16,\ p < .01$), 
while the number of errors in the summaries (manually identified and validated by two researchers) did not differ significantly between conditions ($Z = -0.54,\ p = .58$). 
Subjective ratings further echoed these results: participants found it easier to understand the video’s events with Branch Explorer ($Z = -2.45,\ p <.05$). 
They credited this improved comprehension to the system's ability to encourage re-watching (mentioned nine times) and to support exploration without losing context (six times).

\textbf{Participants highlighted the need for better alignment between narration and frame-level exploration.} 
All participants noted that frame-level spatial exploration often failed to match their interests---for instance, P13 remarked, ``\textit{The narrator mentioned a snow castle, but there's no such object descriptions when I pause.}'' 
This mismatch occurred because users relied on information from the narration, which was not precisely aligned with video frames. 
This finding suggests opportunities to make the system more responsive to the narration context, which we further discuss in Section~\ref{sec:dis_context}.

\subsection{Exploration Strategy (RQ4)}
During the study, participants employed personalized \textbf{strategies} to explore 360° content, often motivated by common \textbf{triggers}. 

\subsubsection{\underline{\textbf{Personalized Exploration Strategy}}} 
\textbf{Most participants preferred lightweight interaction over detailed exploration during first-time viewing.} 
When using Branch Explorer for first-time viewing, participants used phone-shaking to select branches (4.75 times per video, SD = 1.17) more frequently than storyline navigation (0.50 times, SD = 0.74) or spatial exploration (0.42 times, SD = 0.59), because they preferred to ``\textit{get an immersive experience first}'' (P9). 
In contrast, when using the baseline, participants relied more on spatial exploration for first-time viewing (3.21 times, SD = 1.06), as it was ``\textit{the only way to get additional information}'' (P12). 
These findings suggest that lightweight interaction is essential for supporting exploration without disrupting the video flow.

\textbf{Participants responded to vibration notifications based on personal preferences.}
During first viewing, participants interacted with an average of 67.0\% of notifications (SD = 14.8\%). 
For instance, P9 responded to only 35.7\% ``\textit{to get a smooth viewing experience}'', 
while P18 responded to 92.9\% ``\textit{out of curiosity about other choices}''. 
These patterns suggest that the system could adapt to individual preferences---for instance, by proactively presenting options to users with high response rates.

\textbf{Participants showed varied usage patterns of storyline navigation and spatial exploration.} 
With Branch Explorer, eight participants used both features relatively evenly, with each accounting for 30\%–70\% of their interactions. 
They emphasized the distinct roles of each tool: ``\textit{One is for exploring the story, the other for diving into a single moment}'' (P17). 
In contrast, three participants primarily used storyline navigation (>70\%) because they ``\textit{care more about the narrative}'' (P13), while one participant favored spatial exploration (>70\%) to ``\textit{learn more about the environment layout}'' (P9). 
These findings highlight opportunities to offer personalized support---such as suggesting key moments to users who favor spatial exploration.

\subsubsection{\underline{\textbf{Creative Exploration Strategy}}} 
Participants adopted creative strategies during 360° video exploration:

\textbf{(1) Reviewing Story Progression.} 
During replay, nine participants quickly revisited all branching points to ``\textit{review the story}'' (P13). 
They found the scene titles (e.g., ``\textit{[Scene 3 of 7] In the subway}'') particularly useful for skimming the storyline.

\textbf{(2) Deepening Scene Understanding.} 
Twelve participants rapidly switched among branches to ``\textit{fully understand a scene}'' (P12), 
noting that the branch titles (e.g., ``\textit{[Branch 3 of 3] Search for Exits}'') helped them integrate information from multiple viewpoints.

\textbf{(3) Building Spatial Awareness.} 
Ten participants explored paused frames to understand the spatial layout, 
likening the experience to ``\textit{using a white cane}'' (P9). 
They either rotated continuously for an overview, 
or swayed gently to locate specific descriptions.

\textbf{(4) Tracking Temporal Changes.} 
Three participants creatively used spatial exploration to monitor changes in layout over time. For instance, P11 repeatedly paused the video (V1) to track a skier’s movement, describing the experience as ``\textit{the whole world slowed down, giving me time to explore moving objects in detail}''.

\textbf{These findings highlight the potential for novel interaction techniques}---such as slow-motion playback---to better support spatial exploration in temporally changing environments.

\subsubsection{\underline{\textbf{Exploration Triggers}}}\label{sec:triggers}
During the study, participants initiated exploration in response to four primary types of triggers:

\textbf{(1) Anticipated Visual Changes.}
Participants often began exploring in anticipation of visual transitions, aiming to ``\textit{understand a new scene}'' (P12). In addition to vibration notifications, they also relied on semantic cues in the narration---such as descriptions of new settings (P20: ``\textit{You entered a subway station…}'') and announcements of emerging events (P17: ``\textit{An explosion happened}'').

\textbf{(2) Salient Auditory Cues.}
Exploration was triggered when audio cues captured participants' attention. These cues included ``\textit{objects in the narration}'' (P10), audio effects such as ``\textit{cracking sounds}'' (P13), or new speech tones that signaled ``\textit{a new speaker}'' (P16).

\textbf{(3) Narrative Climaxes.}
Participants often explored during narratively intense moments---for instance, \hl{``\textit{at the peak of a fight scene}'' (P12) or ``\textit{during the music chorus}'' (P18).} They initiated exploration to deepen their understanding of these key moments.

\textbf{(4) Conflicting or Missing Information.} 
Participants were prompted to explore further when the audio provided limited or conflicting information---such as during ``\textit{periods without narration}'' (P10) or when ``\textit{the sound did not match narration}'' (P14)---aiming to resolve confusion or fill in informational gaps.

\textbf{These findings suggest opportunities for context-aware descriptions.} For instance, when users initiate exploration, the system could analyze the surrounding video context to predict user intent and surface the most relevant information. We further discuss this direction in Section~\ref{sec:dis_context}.

\section{Discussion}
Our formative study revealed three key considerations for supporting 360° video interaction through branching narratives: 
offering diverse branch options, 
ensuring coherent story progression, 
and enabling immersive navigation across branches. 
To address these needs, Branch Explorer integrates three core modules---branch diversity optimization, 
coherent narration generation, and 
immersive branch navigation---to provide a seamless viewing experience. 
User evaluations showed that Branch Explorer significantly enhanced user agency, immersion, and engagement in 360° video viewing. Participants also developed personalized exploration strategies, highlighting opportunities for more adaptive solutions. 

In the following sections, we discuss lessons learned from our study:
(1) directions for system personalization, 
(2) areas for pipeline improvement, 
(3) implications for accessible visual media, and 
(4) opportunities for context-aware audio descriptions.

\subsection{Directions for System Personalization}\label{sec:personalization}
Based on the evaluation study, we identify directions for tailoring Branch Explorer to different user contexts:

\textbf{Infer User Interest from Interaction History.} 
Several participants showed consistent content preferences across branching points (e.g., P18 repeatedly focused on skiers in V1). 
This suggests that online learning methods \cite{hoi2021online} could help infer user interests and automatically select relevant branches. 
Additionally, three participants proposed an auto-pilot mode that guides users along unvisited paths (P14) or repeats their last selections (P20).

\textbf{Adapt to Personalized Interaction Styles.} 
Participants exhibited personal patterns in using the exploration features, 
indicating opportunities for personalized interaction. 
For example, users who frequently responded to vibration cues (P18) may benefit from auto-presented options, while less responsive users like P9 may prefer fewer notifications to maintain immersion.

\hl{
\textbf{Personalize the Weights of Diversity Metrics.} 
Participants expressed varied preferences for diversity metrics. For example, P15 prioritized social interest, while P16 preferred spatial diversity. 
To accommodate these varied preferences, future systems could allow users to adjust the weights of diversity metrics and even define their own criteria, such as selecting ``\textit{the least viewed path}'' (P13).
}

\hl{
\textbf{Support Acoustic Feature Customization.} 
BLV users have diverse preferences for acoustic features ~\cite{jiang2023beyondAD}, including audio format (e.g., mono vs. spatial presentation), voice tone (e.g., neutral vs. expressive), and narrative perspective (e.g., first-, second-, or third-person). Future systems could support user customization of these acoustic features to enable more immersive and personalized auditory experiences for BLV users.
}

\subsection{Areas for Pipeline Improvement}\label{sec:pipeline_improvement}
We evaluated the system pipeline using 18 videos spanning five genres and a range of visual styles (Section~\ref{sec:technical_eval}). 
In a sample of 100 branch narrations, the system achieved an accuracy of 78\%, with 22 instances containing errors. 
Based on this evaluation and user feedback, we identify several key directions for improvement:

\textbf{(1) Add composition constraints during branch generation.} 
Among the 22 inaccurate narrations, eight errors were due to poor composition, such as key objects appearing near the edges of the viewport. 
This occurs because the pipeline selects viewport centers based on visual saliency peaks, which do not always align with optimal composition or consistently track moving objects \cite{dahou2021atsal}. 
To address this issue, future work could integrate salient object detection \cite{lee2022tracer}, object tracking \cite{zhang2022bytetrack}, and composition-aware constraints \cite{shoot360,tog_360} to guide branches toward better-framed views.

\textbf{(2) Ensure reference consistency across branches.} 
Twelve errors resulted from inconsistent entity references (e.g., referring to two different men as the same person). 
This occurs because the pipeline generates each branch narration using the corresponding video clip and previous branch descriptions, 
which can lead to mismatched references. 
Future work could address this by applying object re-identification techniques \cite{he2021transreid} 
or spatial-temporal video grounding methods \cite{liu2024single} 
to improve entity tracking across branches.

\textbf{(3) Improve narration integration in speech-heavy videos.} 
In speech-heavy videos such as talk shows, the system often fails to find suitable areas for inserting narration. One potential solution is to extend the timeline using looping background music---an approach found acceptable by BLV viewers \cite{amy2020rescribe}---to create space for narration without disrupting the original audio.

\hl{
\textbf{(4) Adapt branching strategies to video content.} 
360° videos vary in the number of regions of interest (ROIs)~\cite{wu2017dataset,liu2023radarvr}. 
To better accommodate this diversity, future branching strategies could improve in two key areas. 
First, regarding \textbf{branch count}, the system's current emphasis on providing diverse options is effective for content with multiple ROIs---such as films and virtual tours, which represent over 75\% of the 360° video market~\cite{marketsizes2024vr}. 
However, simpler content with fewer or singular ROIs (e.g., talk shows with a single viewpoint) may not benefit from such variety. 
Future systems could dynamically adjust branch count based on scene complexity to provide more meaningful choices. 
Second, regarding \textbf{branching timing}, the system currently merges branching points within a 30-second window to reduce interruptions. 
While this approach helps maintain a steady viewing pace, it may miss opportunities to branch at key narrative transitions or visually dense moments. Future strategies could incorporate content-aware cues --- such as story shifts or visual complexity peaks --- to better align branching points with each video's structure.}

\subsection{Implications for Accessible Visual Media}
Our research offers broader implications for: (1) enhancing the accessibility of 2D videos and virtual environments for BLV users, and \hl{(2) improving 360° video accessibility for broader populations.}

\subsubsection{\textbf{2D Video Accessibility for BLV Users}}
\xsc{BLV users have diverse content preferences when consuming 2D videos \cite{asset2024customAD,jiangchi24context}, yet providing such varied information within the limited video time remains challenging \cite{amy2020rescribe,wang2021Tiresias}. 
A promising approach is \textbf{multi-branch descriptions}, where each branch focuses on a specific type of content needed by BLV viewers \cite{jiangchi24context}. 
To enhance user control, future systems could adopt Branch Explorer's interaction design, allowing viewers to flexibly select branches during playback.}

\xsc{Unlike 360° videos, 2D videos have fixed viewports, providing \textbf{new possibilities for branch representation and generation}. 
On the representation side, branches can extend beyond visual descriptions to include richer contextual information---such as relevant background knowledge \cite{whitehead2018incorporating} or real-time user commentary \cite{xu2025danmua11y}. 
On the generation side, techniques like salient object detection \cite{lee2022tracer} and personalized scan-path prediction \cite{jiang2024eyeformer} could help prioritize visual elements, thereby creating diverse descriptions that better address the varied needs of BLV viewers.}

\subsubsection{\textbf{Virtual Environment Accessibility for BLV users}}
\xsc{Compared to 360° videos, virtual environments offer greater freedom of interaction \cite{navstick}---such as walking around and interacting with objects---which can introduce additional cognitive load during exploration \cite{van2013lost}. 
Prior work has explored the use of sighted guides to assist BLV users in navigating virtual spaces \cite{collins2023guide}. 
Future systems could also leverage previous users' interaction data \cite{kepplinger2020see} to identify key actions and moments in virtual environments, 
thereby providing \textbf{interactive walk-throughs} for BLV users. 
Beyond the technical challenge of path identification, 
this direction presents new \textbf{human-factor challenges}, including how to maintain user agency and support spatial awareness during guided exploration \cite{navstick,Surveyor}. 
Additionally, future systems could enhance spatial exploration in dynamic environments by incorporating slow-motion playback, helping BLV users better perceive temporal changes.}

\subsubsection{\textbf{360° Video Accessibility for Broad Populations}}
\hl{Beyond BLV users, branching narratives can also enhance the accessibility of 360° videos for broader populations. For people with \textbf{motor impairments} who may have difficulty moving their head or body \cite{mitchell2024characterizing}, future systems could visually present multiple branches to support motor-free navigation. For \textbf{sighted users}, branching narratives may unlock game-like experiences. Future research could explore methods for optimizing narrative branches for visual consumption, as well as novel interaction and visualization techniques---such as representing branches as portals \cite{portal2}---to further enhance the accessibility and engagement of 360° videos for diverse audiences.}

\subsection{Toward Context-Aware Audio Descriptions}\label{sec:dis_context}
Branch Explorer balances interactivity with narrative continuity using branching narratives. 
Yet, it still involves the \textit{Narrative Paradox} \cite{louchart2003solving}---a tension between user agency and narrative coherence. 
Because the narrations are pre-generated, users cannot make spontaneous choices or fully tailor them to personal interests. 
To address this, we envision \textit{context-aware audio descriptions} that respond dynamically to both video content and viewer interest. 
Based on our study, we outline several key considerations for this approach:

\textbf{First, BLV users primarily seek information relevant to the immediate audio context}, such as objects referenced in narration or special sound effects (Section~\ref{sec:triggers}). 
Future systems could leverage the surrounding audio context to better infer user intent during interactions like menu navigation or playback pauses.

\textbf{Second, paused video frames may not reliably reflect user interest.} Since audio descriptions are not precisely aligned with visual frames \cite{amy2020rescribe}, frame-level descriptions can cause confusion for BLV viewers (e.g., P9: ``\textit{The narration mentioned a car, but I did not find it anywhere.}''). Prioritizing the surrounding audio context over static visuals may lead to more accurate system responses.

\textbf{Third, users benefit from contextual cues when navigating video.} Interruptions such as pausing or skipping can lead to resumption lag \cite{altmann2004task}, where users temporarily forget relevant context due to working memory limitations \cite{baddeley2020working}. Participants in our study expressed a need for brief reminders to help reorient within the narrative (e.g., ``\textit{[Previously] Ground Explosion}'' as used in Branch Explorer). These cues could be made adaptive based on users' viewing history, such as where they paused or skipped.

\hl{\textbf{Fourth, delivering real-time audio descriptions presents both technical and human-factor challenges.} 
On the \textbf{technical} side, audio descriptions should maintain global narrative coherence and integrate smoothly with the original audio track \cite{amy2020rescribe}, which frame-level descriptions alone cannot achieve \cite{uist24worldscribe}. 
Future work could explore generating real-time narration by combining immediate user input, prior narration, and global video context, while carefully managing timing to avoid audio overlap. 
On the \textbf{human} side, future research should explore how to support fluid user input without disrupting narrative continuity. One promising direction is to integrate the \textit{discoverability} of branching narratives with the \textit{flexibility} of visual question answering (VQA) \cite{stangl2023potential}. This integration can further preserve \textit{immersion} by seamlessly embedding VQA responses into the main narration. 
Together, these approaches can enable more interactive and coherent experiences for BLV users.}

\subsection{Limitations and Future Work}
We now summarize the limitations and future work in three aspects: user interaction, branch generation, and system evaluation.

\textit{\textbf{User Interaction.}} 
While this work focuses on textual modalities to enhance 360° video accessibility, future work could explore other modalities to deepen immersion, 
such as spatial audio \cite{jiang2023beyondAD,frontrow}, haptic feedback \cite{Canetroller,virtual_paving,tactile_compass}, or visual augmentation for low vision users \cite{seeingVR,light_guide}. 
\hl{While the system employs a phone-shaking gesture for branch selection, it may shift the on-screen viewport for users with low vision. Future work could explore simple on-screen gestures (e.g., a triple tap) to mitigate this issue.} 
Additionally, while the current system encodes branches as viewing paths, 
their role could be expanded to convey other forms of information useful to BLV viewers, such as relevant background knowledge \cite{whitehead2018incorporating}. 

\textit{\textbf{Branch Generation.}} 
\hl{While the system currently constructs branches using frame-level saliency, future work could explore continuous path prediction~\cite{jiang2024eyeformer} to improve trajectory stability}. 
Real-user viewing data~\cite{jin2022you, heatmap} could also be leveraged to better capture social interest. 
While branching timing is currently determined by video content (e.g., scene transitions), future systems could also identify branching points by analyzing real-user viewing behavior---such as moments when viewing paths diverge significantly. 
The system uses a fully automated pipeline to generate branch narrations, 
which introduces errors due to hallucinations, non-ideal viewport composition, or inconsistent object references. 
These limitations could be mitigated through the improvements discussed in Section~\ref{sec:pipeline_improvement}. 
Future work could also improve description quality using human-in-the-loop strategies---for instance, 
allowing video creators to label branches \cite{amy2017uistguidance}, 
or enabling audio describers to compose branch narrations \cite{chang2022omniscribe}. Additionally, while branch narrations are currently generated offline, future systems could explore real-time description generation in response to user input \cite{uist24worldscribe}.

\textit{\textbf{System Evaluation.}} 
Our user study was conducted with videos under five minutes in length; longer content may introduce new challenges such as viewer fatigue, which needs further investigation. Expanding the evaluation to include a more diverse set of videos also presents a promising direction for future research.
\section{Conclusion}
Branch Explorer leverages branching narratives to enable interactive and immersive 360° video viewing for BLV users. 
It addresses three key considerations---providing diverse branch options, 
ensuring coherent story progression, 
and enabling immersive branch navigation---to make 360° videos more inclusive and engaging. 
User evaluations demonstrated that Branch Explorer significantly enhanced 
BLV users' sense of agency, immersion, and engagement in 360° video viewing. 
We further identified opportunities to personalize the system to accommodate diverse user needs 
and derived broader implications for supporting accessible exploration of visual media. 
We hope this work offers valuable insights into improving 360° video accessibility 
and inspires future research that empowers BLV users to deeply engage with visual content. 

\begin{acks}
The authors would like to thank all participants for their support during the studies. We are also grateful to the reviewers for their constructive feedback. Special thanks to Bingjian Huang, Zichen Liu, Yuying Tang, and Prof. Xian Xu for their insightful discussions. This work is partially supported by the Research Grants Council of the Hong Kong Special Administrative Region under the Hong Kong PhD Fellowship Scheme (No. PF23-94000).
\end{acks}

\bibliographystyle{ACM-Reference-Format}
\bibliography{sample-base}



\end{document}